\documentclass{emulateapj}
\usepackage{color} % For use in the macro \query defined below
\usepackage[colorlinks,urlcolor=blue,citecolor=blue,linkcolor=blue]{hyper ref}
\usepackage{epsfig}
\usepackage{subfigure}
\usepackage{graphicx}
\usepackage{amsmath}
\usepackage{natbib}
\newcommand{\pa}{\partial}
\newcommand{\mb}{\boldsymbol}

\newcommand{\pasa}{Publications of the Astronomical Society of Australia}

\shorttitle{Global Simulations of PPDs}
\shortauthors{X.-N. Bai}

\begin{document}


\title{Global Simulations of the Inner Regions of Protoplanetary Disks with Comprehensive Disk Microphysics}


\author{Xue-Ning Bai}
\affil{Institute for Theory and Computation, Harvard-Smithsonian
Center for Astrophysics, 60 Garden St., MS-51, Cambridge, MA 02138}
\email{xbai@cfa.harvard.edu}


\begin{abstract}
The gas dynamics of weakly ionized protoplanetary disks (PPDs) is largely governed
by the coupling between gas and magnetic fields, described by three non-ideal
magnetohydrodynamical (MHD) effects (Ohmic, Hall, ambipolar). 
Previous local simulations incorporating these processes have revealed that the inner
regions of PPDs are largely laminar accompanied by wind-driven accretion. We conduct
2D axisymmetric, fully global MHD simulations of these regions ($\sim1-20$ AU), taking
into account all non-ideal MHD effects, with tabulated diffusion coefficients and
approximate treatment of external ionization and heating. 
With net vertical field aligned with disk rotation, the Hall-shear instability strongly amplifies
horizontal magnetic field, making the overall dynamics dependent on initial field configuration.
Following disk formation, the disk likely relaxes into an inner zone characterized by
asymmetric field configuration across the midplane that smoothly transitions to a more
symmetric outer zone. Angular momentum transport is driven by both MHD winds and laminar
Maxwell stress, with both accretion and decretion flows present at different heights, and
modestly asymmetric winds from the two disk sides. With anti-aligned field polarity, weakly
magnetized disks settle into an asymmetric field configuration with supersonic accretion flow
concentrated at one side of disk surface, and highly asymmetric winds between the two disk
sides. In all cases, the wind is magneto-thermal in nature characterized by mass loss rate
exceeding the accretion rate. More strongly magnetized disks give more symmetric field
configuration and flow structures. Deeper far-UV penetration
leads to stronger and less stable outflows. Implications for observations and planet formation
are also discussed.
\end{abstract}



\section{Introduction}\label{sec:intro}

Planet formation takes place in protoplanetary disks (PPDs) 
surrounding young stars. Composed of gas and dust, PPDs
offer rich observational diagnostics that help constrain the physical
scenarios of planet formation. With typical lifetime of a few Myrs
\citep{Haisch_etal01}, PPDs are known to be rapidly accreting, with
typical accretion rate of $\sim10^{-8}M_{\odot}$ yr$^{-1}$
(e.g., see \citealp{Hartmann_etal16} for an updated review). The
accretion phenomenon is closely related to jets and outflows that are
ubiquitous among young stellar objects (e.g., see \citealp{Frank_etal14}
for a recent review). The solid materials are primarily probed by the dust
thermal emission at sub-millimeter (mm) as well as in scattered light at
infrared wavelengths (e.g., see \citealp{WilliamsCieza11}). With the
advent of the Atacama Large Millimeter/sub-millimeter Array (ALMA)
and extreme adaptive optics systems such as SPHERE/VLT and the
Gemini Planet Imager, PPDs have revealed rich substructures
(e.g., \citealp{HLTau15,Nomura_etal16,Perez_etal16,Isella_etal16,Ginski_etal16,deBoer_etal16}),
even down to AU scales \citep{Andrews_etal16}.
Furthermore, astrochemistry has emerged to provide more refined
information about the physical environments of PPDs with implications for
planetary composition (e.g.,
\citealp{Oberg_etal11,HenningSemenov13,Qi_etal13,Cleeves_etal16}).

Theoretically, the gas dynamics of PPDs, especially the local and global
disk structure as well as internal flow structure (such as the level of
turbulence), plays a crucial role in almost all
aspects of planet formation \citep{Armitage11}. This is because small dust
grains are coupled with the gas aerodynamically, whereas larger bodies are
coupled with the disk gravitationally. In particular, dust grains always migrate
towards higher pressure, leading to radial drift and particle trapping
at pressure maxima, and these theoretical predictions have found
observational support (e.g., \citealp{Pinilla_etal12,BirnstielAndrews14,Zhang_etal16}).
For planets that form within the disk lifetime, planet-disk interaction leads
to planet migration, and its direction and rate sensitively depend on the
radial gradients of various disk quantities, as well as the level of turbulence
(e.g., see \citealp{Baruteau_etal16} for a recent review).

\subsection[]{Current Understandings of PPD Gas Dynamics}

The central question on PPD gas dynamics lies in the mechanism of angular momentum
transport, which shapes the disk structure and drives disk accretion and global
evolution. Angular momentum can be transported radially (viscous accretion), mainly
mediated by turbulence or large-scale magnetic stress, or vertically, mediated by a
magnetized disk wind
(see \citealp{Turner_etal14} for a recent review). In either scenario, magnetic field
is believed to play an essential role, as we briefly discuss below, and this is further
supported from paleomagnetic studies of the Semarkona chondrite \citep{Fu_etal14}.

In the absence of magnetic fields, a number of mechanisms have been studied,
such as the vertical-shear instability \citep{Nelson_etal13,StollKley14,LinYoudin15},
the convective overstability \citep{LesurPap10,LyraKlahr11,KlahrHubbard14,Lyra14},
and the zombie vortex instability \citep{Marcus_etal13,Marcus_etal15}. Nevertheless,
besides the fact that these instabilities all require certain thermodynamic
conditions to operate, the resulting level of turbulence is typically weak, with
Shakura-Sunyaev $\alpha$ reaching at most $\sim10^{-3}$, which is too small
to account for the accretion rates in the bulk disk population (e.g.
\citealp{Andrews_etal09,Andrews_etal10}). Other mechanisms, such
as the gravitational instability \citep{Gammie01,Rafikov09}, and spiral-density
waves driven by envelop infall \citep{Lesur_etal15}, can provide significant angular
momentum transport, although only at the very early stages of PPD evolution.

With magnetic fields, the key microphysical processes involve determining how well
magnetic fields are coupled with the gas. The ionization of PPDs largely relies
on external sources such as cosmic-rays and X-rays, leading to extremely low level
of disk ionization with vertically stratified ionization structure
\citep{Sano_etal00,IlgnerNelson06,BaiGoodman09}. 
As a result, magnetic fields are no longer frozen in to the bulk gas as in ideal
magnetohydrodynamics (MHD), introducing three non-ideal MHD effects: Ohmic
resistivity, the Hall effect and ambipolar diffusion (AD).
The three effects control the gas dynamics in different ways, and at fixed field strength,
the dominant effect transitions from resistivity to the Hall effect, and to AD as density
decreases \citep{Wardle07,Bai11a}.

Conventionally, the magnetorotational instability (MRI, \citealp{BH91}) has been
considered as the dominant mechanism to drive disk accretion. However, the
MRI is strongly affected by non-ideal MHD effects. Linear modes are
damped by resistivity and AD \citep{BlaesBalbus94,Jin96,KunzBalbus04}, 
whereas the Hall effect modifies the dispersion relation depending on the polarity
of vertical field threading the disk \citep{Wardle99,BalbusTerquem01}.
Taking only Ohmic resistivity into account, the picture of layered accretion has been
established \citep{Gammie96,FlemingStone03,TurnerSano08,OishiMacLow09}, where
the MRI is suppressed by resistivity in the midplane region of the inner disk
(i.e., the densest region where resistivity dominates), while it operates in the much
better ionized, low-density disk surface, driving viscous accretion through the surface
layer. 

The conventional picture of layered accretion no longer holds when AD is further
taken into account. As AD dominates towards low density regions, the MRI is found
to be almost entirely suppressed in the inner region of PPDs ($\lesssim15$ AU,
\citealp{BaiStone13b,Bai13,Gressel_etal15}). The MRI is substantially damped in
the low-density outer disk \citep{Simon_etal13a,Simon_etal13b,Bai15}, whereas it can,
however,
operate in the disk surface thanks to far-UV (FUV) ionization \citep{PerezBeckerChiang11b}.
In other words, layered accretion is likely more applicable to the outer instead of the inner disk.
As the MRI is suppressed or damped, efficient angular momentum transport requires
the disk to be threaded with net vertical magnetic field, and a magnetized disk wind is
likely the primary mechanism to drive disk accretion. 

The inclusion of the Hall effect further makes the gas dynamic depend on the polarity
of the net vertical field. In particular, when vertical field is aligned with the disk rotation
axis, horizontal components of the field are amplified due to the Hall-shear instability
\citep{Kunz08}, which leads to enhanced radial transport of angular momentum by
large-scale magnetic stresses \citep{Lesur_etal14,Bai14,Bai15,Simon_etal15b}. In the opposite
case of anti-aligned vertical field, horizontal field is reduced towards zero. In both cases,
disk winds likely remain the dominant mechanism to drive disk accretion \citep{Bai14}.
In the outer disk where the MRI is damped, the Hall effect can enhance/reduce
turbulence depending on polarity \citep{SanoStone02a,SanoStone02b,Bai15,Simon_etal15b},
though not substantially.

MHD disk winds have been studied extensively in the literature, ranging from global
self-similar solutions (e.g., \citealp{BlandfordPayne82,Li95,Ferreira97})
to local solutions that match to such solutions (e.g.,
\citealp{WardleKoenigl93,Konigl_etal10,Salmeron_etal11}),
to global simulations (e.g.,
\citealp{Krasnopolsky_etal99,Pudritz_etal06,Zanni_etal07,Tzeferacos_etal09}).
The wind properties are now well known to depend mainly on the strength and
distribution of the magnetic flux threading the disk, and on the mass loading, with the
latter mainly being controlled by disk physics and thermodynamics.
We emphasize that previous studies are not directly applicable to PPDs because
the main disk microphysics (all three non-ideal MHD effects with realistic ionization
structure) and thermodynamics were not properly taken into account. Most
studies considered vertical field strength near equipartition (to avoid the development
of the MRI), and the resulting wind-driven
accretion rate would be orders of magnitude larger than typical PPD accretion rates.
On the other hand, realistic local simulations have demonstrated that a weak vertical
field can naturally sustain wind launching, and drives disk accretion at desired accretion
rates \citep{BaiStone13b,Bai13}.

\subsection[]{Outstanding Issues}\label{ssec:issues}

These recent works have revealed rich disk physics resulting from the non-ideal MHD
effects, highlighting the importance of incorporating realistic disk microphysics for studying
PPD gas dynamics. However, the aforementioned works are mostly local simulations. One
exception is \citet{Gressel_etal15}, where the simulations were radially global, yet
vertically local, and the Hall effect was not included. Another exception is the very recent
work of \citet{Bethune_etal17}, which we will discuss in more detail in Section
\ref{ssec:compare}.
Three outstanding issues remain to be worked out and clarified.

First, wind kinematics. Local simulations fail to cover the full depth of the gravitational
potential well of the central star, and the wind mass loss rate has been found to depend
on the vertical height of the simulation box \citep{Fromang_etal13,BaiStone13b}.
This effect can be understood from a different point of view: the wind is artificially
truncated by the imposed boundary conditions. To overcome this limitation, the
computational domain must be sufficiently extended so that wind flow passes major
critical points (i.e., wind velocity exceeds sonic/Alfv\'en speed) and loses causal
connection with the disk surface.

In anticipation of this work, we developed a semi-analytical theory for MHD disk
winds from PPDs in \citet{Bai_etal16} that allows the flows to pass all critical points to arrive
at unique solutions. We have also adopted an approximate treatment of the
thermodynamics of the disk wind to mimic FUV/X-ray heating, which has conventionally
been considered to drive photoevaporation (as a pure thermal wind, see
\citealp{Alexander_etal14} for a review). The wind solutions, which we call
magneto-thermal disk winds, indicate that PPDs lose mass from disk winds at a rate
comparable to wind-driven accretion rate, and wind kinematics is most sensitive to poloidal
field strength as well as how deep FUV/X-ray can penetrate into (and hence heat and
ionize) the disk. These predictions remain to be verified and calibrated through
realistic global simulations.

Second, the symmetry issue. A physical wind geometry requires that poloidal field lines
bend away from the central star, which further requires that toroidal field must change
sign across the disk. The toroidal field gradient is also directly associated with the
torque exerted by the disk wind, which drives the accretion flow. In \citet{BaiStone13b}
and \citet{Bai13}, we found in local shearing-box simulations that in the inner disk, the
flip of toroidal field occurs at a location that is offset from the midplane, resulting in
symmetry breaking, which is later confirmed in global simulations \citep{Gressel_etal15}.
Including the Hall effect with aligned vertical field, the Hall-shear instability
amplifies the toroidal field near the midplane so strongly that in local simulations,
toroidal field of one sign overwhelms and no flip can be sustained in the
simulation box \citep{Bai14}. While we speculated that this is an artifact of vertical
boundary condition, global simulations are essential to resolve this issue. This is also
crucial to determine the global field configuration and flow structures in PPDs.

Third, the origin and evolution of magnetic flux. As discussed earlier, the presence of
net poloidal magnetic flux is essential to drive disk accretion. We further showed more
quantitatively that global disk evolution is primarily governed by the amount of flux
threading the disks, and its radial distribution \citep{Bai16}. Therefore, transport of
magnetic flux in PPDs is an even more fundamental question.
In \citet{BaiStone17}, we conducted preliminary studies of magnetic flux transport in
PPDs, and emphasized the unique roles played by the Hall effect and AD that has
been overlooked in the literature
\citep{Lubow_etal94a,Okuzumi_etal14,GuiletOgilvie14}. We found that when the disk
is laminar, magnetic flux is systematically transported outward in a polarity-dependent
manner, with rate of transport being faster in the anti-aligned case. We also noted that the
exact rate of transport can be sensitive to the details of the disk ionization structure,
which we treated very roughly, and more realistic simulations are needed to
quantitatively characterize the rate of magnetic flux transport in PPDs.

\subsection[]{This Work}

We aim to conduct global simulations of PPDs that incorporate the most realistic
disk microphysics. A unique aspect of our simulations is that we properly resolve the
most important disk microphysics, and in the mean time the computational domain
extends all the way to near the polar region, which is essential to accommodate the
launching and propagation of MHD disk winds, as well as to accommodate magnetic
flux evolution. 
Our simulations include all three non-ideal MHD effects with realistic
treatment of disk ionization chemistry, together with approximate treatment of disk
thermodynamics. These simulations are made possible thanks to the newly
developed Athena++ MHD code (Stone et al. in preparation).
We have implemented all non-ideal MHD effects \citep{BaiStone17}, and carefully
designed simulation setup that circumvent difficulties especially associated with
boundary conditions. With these simulations, we are able to address the aforementioned
three major issues simultaneously, offering the most realistic PPD simulations to date.

This paper is organized as follows. In Section \ref{sec:setup}, we provide detailed
descriptions of the numerical method and simulation setup. Main diagnostics are
discussed in Section \ref{sec:diag}. We focus on three fiducial simulations and
analyze the results of each simulation in detail in Sections \ref{sec:nohall}-\ref{sec:antialign}. 
A brief parameter study is conducted in Section \ref{sec:param}.
The results are discussed in broader contexts in Section \ref{sec:discussion}, and
we summarize and conclude in Section \ref{sec:sum}.

\section[]{Method}\label{sec:setup}

We use the newly developed grid-based higher-order Godunov MHD
code Athena++ (Stone et al., in preparation), which is the successor of the widely
used Athena MHD code \citep{GardinerStone05,GardinerStone08,Stone_etal08}, to
carry out global simulations of PPDs in this work. Athena++ works for curvilinear
coordinate systems (here we use spherical-polar coordinates), where geometric source
terms are properly implemented that guarantees angular momentum conservation. It
also employs flexible grid spacings, allowing simulations to be performed over large
dynamical ranges. All three non-ideal MHD terms have been implemented
\citep{BaiStone17}, which are the key microphysical processes for our simulations.

\subsection[]{Dynamical Equations}

Using Athena++, we solve the MHD equations in conservative form, including
non-ideal MHD effects
\begin{equation}
\frac{\pa\rho}{\pa t}+\nabla\cdot(\rho{\mb v})=0\ ,
\end{equation}
\begin{equation}
\frac{\pa(\rho{\mb v})}{\pa t}+\nabla\cdot\bigg(\rho{\mb v}{\mb v}
-\frac{{\mb B}{\mb B}}{4\pi}+{\mathsf P}^*\bigg)=-\nabla\Phi\ ,
\end{equation}
\begin{equation}\label{eq:ind}
\frac{\pa{\mb B}}{\pa t}=\nabla\times({\mb v}\times{\mb B})-
\frac{4\pi}{c}\nabla\times(\eta_O{\mb J}+\eta_H{\mb J}\times{\mb b}+\eta_A{\mb J}_\perp)\ ,
\end{equation}
\begin{equation}\label{eq:eng}
\frac{\pa E}{\pa t}+\nabla\cdot\bigg[(E+P^*){\mb v}
-\frac{{\mb B}({\mb B}\cdot{\mb v})}{4\pi}\bigg]=-\Lambda_c\ ,
\end{equation}
where $\rho$, ${\mb v}$, and $P$ are gas density, velocity and pressure,
$P^*=P+B^2/8\pi$ is total pressure, ${\mathsf P}^*\equiv P^*{\mathsf I}$ with
${\mathsf I}$ being the identity tensor, ${\mb B}$ is magnetic field, with
${\mb b}\equiv{\mb B}/B$ being the unit vector along the field direction.
Total energy density is given by $E=P/(\gamma-1)+\rho v^2/2+B^2/8\pi$,
where $\gamma$ is the adiabatic index, and ${\mb J}$ is the current density,
with ${\mb J}_\perp=-({\mb J}\times{\mb b})\times{\mb b}$ being the component of
${\mb J}$ that is perpendicular to the magnetic field.

We specify the static gravitational potential of the protostar as $\Phi=-GM_*/r$, with
$M_*=M_{\odot}$. Magnetic diffusivities are represented by $\eta_O, \eta_H$ and
$\eta_A$ for Ohmic resistivity, the Hall effect and ambiploar diffusion (AD), which
depend on ionization chemistry to be described in Section \ref{ssec:chem}.
Thermodynamics is mainly controlled by a thermal relaxation term $\Lambda_c$ in
Equation (\ref{eq:eng}).
Note that in this equation, we have not included the Poynting flux (and hence heating)
from non-ideal MHD terms. While not fully self-consistent, ignoring this contribution is mainly
for convenience because we only treat disk thermodynamics approximately by artificially
relaxing disk temperature to a target temperature in relatively short timescales through the
$\Lambda_c$ term. This approach renders the discrepancy largely irrelevant.
More details will be given in Section \ref{ssec:thermo}.

The above equations are written in c.g.s. units, which will be used consistently
in this paper. In the simulations, factors of $4\pi$ are absorbed into the definition
of $B$ so that magnetic permeability is $\mu_B=1$.
The equations are solved in spherical-polar coordinates $(r, \theta)$ in 2D with
axisymmetry. For convenience, we also use cylindrical coordinates $(R, z)$ in this
work to facilitate analysis.

\subsection[]{Basic Disk Model}

Motivated from observations of PPDs (e.g., \citealp{Andrews_etal09,Andrews_etal10}),
we consider a power-law disk surface density profile with exponential cutoff
\begin{equation}
\Sigma(R)=\Sigma_0\ R_{\rm AU}^{-q_S}\exp(-R/R_c)\ {\rm g\ cm}^{-2}\ ,
\end{equation}
where $R_{\rm AU}$ is radius normalized to $1$ AU, $R_c$ is the outer
radius of the disk beyond which the disk surface density cuts off. We take
$\Sigma_0=500$ g cm$^{-2}$ with power-law index $q_S=1$.
We are mainly interested in the inner regions of PPDs with $R\sim1-20$ AU,
where the disk is expected to be largely laminar \citep{Bai13}, and simply set
$R_c=30$ AU.
We note that while $\Sigma$ should be defined by
integrating gas density over the vertical column, in practice, we calculate
$\Sigma$ by integrating over $\theta$ at fixed $r=R$ for convenience: it makes
little difference as long as the disk is thin.

The temperature of the bulk disk is taken to follow the minimum-mass solar
nebular scaling (MMSN, \citealp{Weidenschilling77b,Hayashi81})
\begin{equation}
T(R)=T_0R_{\rm AU}^{-q_T}\ ,\label{eq:tdisk}
\end{equation}
where we take the power-law index $q_T=1/2$.
Disk temperature sets the isothermal sound speed, given by
$c_s(R)^2=P/\rho=k_BT(R)/\mu m_p$, where $\mu=2.34$ is the mean molecular
weight in the bulk disk (molecular gas), $k_B$ is the Boltzmann constant, $m_p$
is proton mass. The disk scale height is given by $H_d(R)=c_s(R)/\Omega_K(R)$,
where $\Omega_K(R)=\sqrt{GM/R^3}$ is the Keplerian angular velocity. 
With $q_T=1/2$, the disk is flared, with disk aspect ratio $\epsilon_d$ given by
\begin{equation}
\epsilon_d(R)\equiv\frac{H_d(R)}{R}\propto R^{(1-q_T)/2}=R^{1/4}\ .\label{eq:hord}
\end{equation}
For an MMSN disk, we have $T_0=280$K and $\epsilon(R=1{\rm AU})=0.034$.
In our simulations, we slightly enlarge the disk thickness with
$\epsilon_d(R=1{\rm AU})=0.045$ in order to adequately resolve the disk with
available computational resources (see Section \ref{ssec:setup} for details).

Assuming disk temperature is vertically isothermal, the gas density is given by
solving the vertical hydrostatic equilibrium
\begin{equation}\label{eq:rho}
\rho(R,z)%=\rho_{\rm mid}(R)\exp({-z^2/2H^2})\\
%=\rho_0\bigg(\frac{r}{R_0}\bigg)^{-\frac{9}{4}}
=\rho_{\rm mid}(R)\exp{\bigg[
%\frac{GM}{c_s(R)^2}\bigg(\frac{1}{r}-\frac{1}{R}\bigg)\bigg]}\ ,
\frac{1}{\epsilon_d(R)^2}\bigg(\frac{R}{r}-1\bigg)\bigg]}\ ,
\end{equation}
where $r=\sqrt{R^2+z^2}$, and
%$\rho_{\rm mid}(R)=\Sigma/\sqrt{2\pi}H\propto R^{-q_s+(q_T-3)/2}=R^{-9/4}$.
$\rho_{\rm mid}(R)=\Sigma/\sqrt{2\pi}H\propto R^{-q_D}$, with
$q_D\equiv q_S+(3-q_T)/2=9/4$.

In the radial direction, radial pressure gradient modifies the rotation
profile to yield\footnote{We have ignored the exponential surface density cutoff
at large $R$ in deriving this equation.}
\begin{equation}\label{eq:vphi}
v_\phi(R,z)=v_K(R)\sqrt{1-(q_S+q_D)
\bigg(\frac{H}{R}\bigg)^2+q_T\frac{R-r}{r}}\ ,
\end{equation}
where $v_K(R)=R\Omega_K(R)$ is the Keplerian speed. 

\subsection[]{Thermodynamics}\label{ssec:thermo}

The bulk of the PPD is heated by the thermal radiation from the protostar,
with temperature approximately given by to Equation (\ref{eq:tdisk}).
The bulk disk is connected to a tenuous disk atmosphere, which is subject to
significant heating from higher energy radiation, especially far-UV (FUV) and
X-rays, and reaches much higher temperatures (e.g.,
\citealp{Glassgold_etal04,Walsh_etal12}). The thermodynamics of the disk
atmosphere is much more complex, and detailed modeling would involve
UV and X-ray radiative transfer calculations coupled with photochemistry
(e.g. \citealp{Woitke_etal16,Haworth_etal16}). While there are also significant
uncertainties, particularly related to the unknown abundance of very small grains,
the general result is that there is a rapid temperature transition as high-energy
photons are exhausted.

Based on the above, we adopt an approach following the simplified treatment
in semi-analytical models of \citet{Bai_etal16} and \citet{Bai16} (see Figure 1
in these two papers). The disk is divided into a cold {\it disk zone}, and a warm
{\it atmosphere}, where the division is set by the depth that high-energy radiation
(especially FUV) is able to penetrate into the disk. Let $\Sigma_{\rm FUV}$ be
this penetration depth (in units of column density), which is treated as a simulation
parameter and is assumed to be a constant. In the simulations, we trace radial
rays from the central star in spherical grid to obtain the column densities
traversed by the rays, denoted by $\Sigma_r(r, \theta)$. For convenience, we
express gas temperature in terms of the local disk aspect ratio
$\epsilon\equiv H/R$, with
\begin{equation}
T(r,\theta)\propto\frac{P}{\rho}=[\epsilon(r,\theta)v_K(R)]^2\ .\label{eq:Tepsi}
\end{equation}
We assume that within the penetration column of FUV radiation, the disk is heated
to a temperature that corresponds to a constant local
$\epsilon=\epsilon_a\equiv0.2$.\footnote{In \citet{Bai_etal16}, we find that the wind
properties is sensitive to $\Sigma_{\rm FUV}$, whereas the detailed temperature
structure in the wind zone is of less importance.}
Beyond this column density, the disk aspect ratio returns to $\epsilon_d$ given by
(\ref{eq:hord}). We then prescribe a smooth but rapid transition to join the two
limits, given by
\begin{equation}\label{eq:fullepsi}
\epsilon(r,\theta)=\frac{\epsilon_d+\epsilon_a}{2}+\frac{\epsilon_a-\epsilon_d}{2}
\tanh\bigg[3\log\bigg(\frac{\Sigma_{\rm FUV}}{\Sigma_r(r,\theta)}\bigg)\bigg]\ .
\end{equation}
To some extent, this treatment is a realization of the semi-analytical global disk evolution
model of \citet{Bai16}.

Equations (\ref{eq:Tepsi}) and (\ref{eq:fullepsi}) define the target disk temperature
$T_c(r, \theta)$. In the simulations, we relax the temperature to $T_c$ with a simple
cooling prescription
\begin{equation}\label{eq:cooling}
\Lambda=-\frac{T-T_c}{\tau}\ ,
\end{equation}
where the relaxation time $\tau$ is set by
\begin{equation}
\tau^{-1}=\Omega_K(R)\bigg(1+
\frac{\Sigma_{\rm FUV}}{\max{[\Sigma_r(r,\theta), 0.1\Sigma_{\rm FUV}]}}\bigg)\ .
\end{equation}
Basically, we relax the gas in orbital timescale in the disk zone. This prescription
avoids the development of vertical shear instability, which requires fast
cooling \citep{Nelson_etal13}. The relaxation time is gradually reduced to about
$1/10$ orbital time in the atmosphere, which we find is necessary to sufficiently
heat the gas to approach $T_0$.
%\citet{LinYoudin15}

We take the adiabatic index $\gamma=5/3$ in our simulations. This is more
appropriate for the atomic gas in the atmosphere/wind zone. In reality, the gas in
the system transitions to become largely molecular in the bulk disk, leading to an
abrupt change in $\gamma$ at the disk surface. As a caveat, this may result in
additional temperature variations in the transition region, and may affect wind
kinematics near the wind base. While this is not captured in our treatment, thermal-chemical
calculations in hydrostatic disks generally found monotonic increase of temperature
from disk to atmosphere (e.g., \citep{Walsh_etal12}). Moreover, we do not find
appreciable difference in the disk dynamics by using a constant $\gamma=7/5$.

Overall, thermodynamics is treated very approximately in several aspects mentioned
previously. In this work, we mainly focus on the role of non-ideal MHD effects on
the overall gas dynamics (see the next subsection), which likely play a dominant role
governing the overall disk angular momentum transport and flow structure. The
simplified treatment of thermodynamics allows us explore these aspects in a more
controlled manner. On the other hand,
we expect that our treatment of thermodynamics at least captures the most essential
ingredients, with $\Sigma_{\rm FUV}$ being the main controlling parameter.

\subsection[]{Ionization, Chemistry and Non-ideal MHD Effects}\label{ssec:chem}

The strength of non-ideal MHD effects is determined by the ionization degree,
or more precisely by the abundance of all charge carriers, a result of disk
chemistry initiated by the ionization processes. We focus on regions not too
close to the protostar so that the disk is not sufficiently hot to trigger thermal
ionization ($\lesssim800$K, \citealp{DeschTurner15}). Below we discuss the
non-thermal ionization processes and the calculation of the magnetic
diffusivities.

\subsubsection[]{Ionization Rates}

The main ionization sources in the bulk disk include cosmic-rays, X-rays.
We follow standard prescriptions, where the ionization rates are given as a
function of column densities. In addition to tracing radial rays to obtain
$\Sigma_r(r,\theta)$, we further trace $\theta-$rays from the upper and lower
poles towards the disk at constant $r$, and define two column densities
$\Sigma_\theta^{\rm top}(r,\theta)$ and $\Sigma_\theta^{\rm bot}(r,\theta)$.
While these rays are not straight, the two column densities reach physically
meaningful values only close to the bulk disk, where the rays are largely vertical
in geometrically thin disks considered here.

For cosmic-ray ionization, the ionization rate is given by \citep{UN81}
\begin{equation}
\zeta_{\rm CR}=10^{-17}\bigg[
\exp\bigg(-\frac{\Sigma_\theta^{\rm top}}{\Sigma_{\rm CR}}\bigg)
+\exp\bigg(-\frac{\Sigma_\theta^{\rm bot}}{\Sigma_{\rm CR}}\bigg)\bigg]
\ {\rm s}^{-1}\ .
\end{equation}
where $\Sigma_{\rm CR}=96$ g cm$^{-2}$. We note that the rate of cosmic-ray
ionization rate bears large uncertainties (e.g., \citealp{McCall_etal03,Cleeves_etal13}),
though it would not change the fact that the midplane region of the inner disk is
extremely weakly ionized, suppressing the MRI \citep{Bai11a}.

For X-ray ionization, we adopt the fitting formula of \citet{BaiGoodman09},
based on calculations of \citet{IG99}. We assume an X-ray luminosity of
$L_X=10^{30}$erg s$^{-1}$, and use the fitting coefficients at X-ray
temperature $T_X=3$ keV, which gives
\begin{equation}
\begin{split}
\zeta_{\rm X}=&\bigg(\frac{r}{1{\rm AU}}\bigg)^{-2.2}\frac{L_X}{10^{30}{\rm erg\ s}^{-1}}
\bigg\{\zeta_1e^{-(\Sigma_r/5\Sigma_{X,a})^{\alpha}}\\
+&\zeta_2[e^{-(\Sigma_\theta^{\rm top}/\Sigma_{X,s})^{\beta}}
+e^{-(\Sigma_\theta^{\rm bot}/\Sigma_{X,s})^{\beta}}]\bigg\}\ .
\end{split}
\end{equation}
The first term accounts for the direct absorption of the X-rays along radial
rays, with $\zeta_1=6.0\times10^{-11}$s$^{-1}$,
$\Sigma_{X,a}=3.6\times10^{-3}$g cm$^{-2}$, and $\alpha=0.4$.
We note that for the direct absorption component, the $\Sigma_{X,a}$ value in
the original fitting formula corresponds to the vertical instead of radial column
density. We here multiply $\Sigma_{X,a}$ by a geometric factor of $5$ to
account for the conversion.\footnote{This is a very approximate
estimate, depending on the disk thickness and the level of disk flaring, etc.
\citet{PerezBeckerChiang11b} quoted a a factor of $\sim3$, though for the very
thin inner PPD, the factor is likely larger.}
The second term describes the scattered X-rays that penetrate deeper,
with $\zeta_2=1.0\times10^{-14}$s$^{-1}$, $\Sigma_{X,s}=1.7$g cm$^{-2}$,
$\beta=0.65$. 

In addition, we include ionization from the decay of short-lived radionuclides
by adding a constant ionization rate of
\begin{equation}
\xi_{\rm SLR}=6.0\times10^{-19}{\rm s}^{-1}\ .
\end{equation}
Note that the rate is expected to be in the range of
$\sim(1-10)\times10^{-19}$s$^{-1}$ and is higher at smaller radii and early times
\citep{UmebayashiNakano09,Cleeves_etal13b}.

\subsubsection[]{Disk Chemistry and Diffusivity Table}

In the bulk disk, ionization-recombination equilibrium is typically achieved within
dynamical time \citep{Bai11a}, especially in the presence of grains. In this case,
the ionization degree, and hence magnetic diffusivities,
are functions of the ionization rate discussed in the previous subsection, as well
as gas density and temperature. We make magnetic diffusivity tables from
chemistry calculations described below.

We use a complex chemical reaction network
developed in \citet{BaiGoodman09} and \citet{Bai11a}, based on the work of
\citet{IlgnerNelson06}, with $175$ gas-phase species. Since \citet{Bai14}, we
extract gas-phase reactions from 2012 UMIST database described in
\citet{UMIST12}, adopting the updated rate coefficients from the new database.
In total, there are 2147 gas-phase reactions, including four ionization reactions
and one ad hoc reaction to account for H$_2$ formation.\footnote{We have
made two changes in the chemistry calculations compared with our previous
work (e.g., \citealp{Bai11a,Bai14}). First, in case when temperature falls out of
the range of validity of the fitting formula provided in in the data base, we now
still use the fitting formula as if it remained valid. See Section 3.1 of
\citet{XuBai16} for further explanation. Second, for electron-grain collisions, we
now adopt a constant sticking coefficient of $s_e=0.3$, instead of directly calculating
$s_e$ described in the Appendix of \citet{Bai11a}. This change follows from
\citet{Ivlev_etal16}, see also \citet{WeingartnerDraine01}.}
In addition, a population of single-sized grains are also included in the network,
with maximum grain charge set to $\pm3$. We choose
grain size $a=0.1\mu$m and mass fraction of $f=10^{-4}$, where the total surface
area is comparable to those obtained by more realistic grain
coagulation/fragmentation calculations \citep{Birnstiel_etal11}.
For a given set of parameters, we start from single-element species and
evolve the network species for $3\times10^6$ years, which is sufficient to reach
chemical equilibrium over a wide range of parameter space. Even the abundance
of some species still show secular evolution trends, the ionization fraction
converges well before the end of the calculation.

Magnetic diffusivities are calculated following standard formulas
\citep{Wardle07,Bai11a}. Ohmic resistivity is always independent
of magnetic field strength. For Hall and ambipolar diffusivities, we have
$\eta_H\propto B$ and $\eta_A\propto B^2$ when field is either weak or
very strong. Complex dependence on $B$ may be present at intermediate
field strength in the presence of charged grains \citep{XuBai16}.
Nevertheless, as we studied in detail in \citet{XuBai16}, unless small grains
are very abundant (say $0.1\mu$m grains with $f=10^{-2}$), the weak field
limit is always satisfied in practice.
Therefore, it suffices to assume $\eta_O$, $\eta_H/B$ and $\eta_A/B^2$,
as we adopt in the table.

Further complications in determining the magnetic diffusivities may arise from
the non-linear Ohm's law \citep{OkuzumiInutsuka15,MoriOkuzumi16} that we
have not accounted for.
Nevertheless, the theory has only been worked out for Ohmic resistivity,
which operates in the presence of very strong current, while it is less likely
to be relevant in disks that are largely laminar, as we have in this work.

\subsubsection[]{FUV Ionization in the Disk Atmosphere}

High-energy radiation, not only heat the disk atmosphere,
but also significantly boost its ionization level. In particular, FUV can
fully ionize atomic carbon and sulfer, raising the ionization fraction to
$x_e\equiv(n_e/n)=10^{-5}$ to $10^{-4}$ \citep{PerezBeckerChiang11b},
with a sharp transition in $x_e$ at the FUV ionization front. Here, $n_e$
and $n$ are the number densities of the electrons and neutrals,
respectively. Note that our chemical network does not include
photo-reactions to account for FUV ionization, but we mimic this effect by
setting the ionization fraction due to FUV to be
\begin{equation}\label{eq:xeFUV}
x_{e, {\rm FUV}}=2.0\times10^{-5}
\exp{\bigg[-\bigg(\frac{\Sigma_r(r,\theta)}{\Sigma_{\rm FUV}}\bigg)^4\bigg]}\ .
\end{equation}
The resulting non-ideal MHD diffusion coefficients are evaluated
according to (applicable when $x_e\gg$ charged grain abundance)
\begin{equation}
\begin{split}\label{eq:etaFUV}
\eta_O&=\frac{c^2m_e}{4\pi e^2}x_e^{-1}\langle\sigma v\rangle_e\ ,\quad
\eta_H=\frac{cB}{4\pi en}x_e^{-1}\ ,\\
\eta_A&=\frac{B^2}{4\pi\langle\sigma v\rangle_im_nn^2}x_e^{-1}\ ,
\end{split}
\end{equation}
where $m_n=\mu m_H$ is the mean molecular mass,
$\langle\sigma v\rangle_e\approx8.3\times10^{-9}(T/100{\rm K})^{1/2}$cm$^3$ s$^{-1}$,
and $\langle\sigma v\rangle_i\approx2.0\times10^{-9}$cm$^3$ s$^{-1}$
are coefficients of momentum exchange in electron-neutral and ion-neutral collisions
\citep{Draine11}. We calculate the diffusivities both from the diffusivity
table, as well as from (\ref{eq:etaFUV}) with $x_e$ given by (\ref{eq:xeFUV}),
and set the diffusivities to be the ones with smaller values, which guarantees
a smooth transition from the disk zone to the atmosphere.

In practice, we further boost $x_{e, {\rm FUV}}$ by a factor $g$ given by
\begin{equation}
g\equiv\exp{\bigg[\frac{0.3\Sigma_{\rm FUV}}
{\Sigma_r(r,\theta)+0.03\Sigma_{\rm FUV}}\bigg]}\ .
\end{equation}
The sole purpose of this factor is to ensure that the gas behave in the ideal
MHD regime throughout the wind zone (otherwise AD would become
progressively more important, while this is not the case from more realistic
calculations as shown in Figure 9 of \citealp{Walsh_etal12}).

The value of $\Sigma_{\rm FUV}$ is uncertain, and is particularly sensitive to the
abundance of very small grains. \citet{PerezBeckerChiang11b} found
$\Sigma_{\rm FUV}\sim0.01-0.1$ g cm$^{-2}$ in their 1D calculations. The
value they quote has been converted to a vertical column density with a
geometric factor of $0.3$. Since we attenuate FUV along radial rays, we
expect a range of $\Sigma_{\rm FUV}\sim0.03-0.3$ g cm$^{-2}$ to be more
appropriate, and we take $\Sigma_{\rm FUV}=0.03$ g cm$^{-2}$ as fiducial.

\subsection[]{Simulation Setup}\label{ssec:setup}

We set the radial grid to span from $r_{\rm in}=0.6$ AU to $r_{\rm out}=60$ AU
with logarithmic grid spacing. The $\theta-$grid extends from the midplane all the
way to near the poles (leaving only a $2^\circ$ cone at each pole), which we find
is essential to properly accommodate the MHD disk wind. The $\theta-$grid is
concentrated around the disk which guarantees adequate resolution, where
$\Delta\theta$ increases by a constant factor per cell from midplane to pole, with
a contrasting factor of $3.5$ between minimum and maximum $\Delta\theta$. The
full grid size is $1152\times512$ in ($r, \theta$), so that at $R=3$ AU where disk
aspect ratio $\epsilon_d\approx0.06$, we achieve a grid resolution of $15$ cells
per $H_d$ in $r$, and $20$ cells per $H_d$ in $\theta$.

We initialize the disk from the hydrostatic solution (\ref{eq:rho}) and (\ref{eq:vphi}),
where disk temperature is set to be vertically isothermal with
$\epsilon(r,\theta)=\epsilon_d(R)$.
A density floor of $\rho(r)=10^{-8}\rho_{\rm mid}(r)$ is set to prevent excessive
density drop in the disk atmosphere in the initial condition. This density floor is
sufficiently small so that upon achieving quasi-steady state, the density in the
disk atmosphere is well above the floor value thanks to disk wind launching.

In the simulations, we trace radial rays and $\theta-$rays to compute
$\Sigma_r(r,\theta)$ and $\Sigma_\theta^{\rm top, bot}(r,\theta)$.
This operation involves global communications, and is executed only at a
time interval of $0.5\Omega_K^{-1}(r_{\rm in})$ (corresponding to at least a few
thousand timesteps) so that it does not affect the overall code performance.

We first run the simulations for $300\Omega_K^{-1}(r_{\rm in})$ without magnetic
fields, which allows the disk atmosphere to be heated to desired temperature
according to (\ref{eq:fullepsi}), and the gas density to adjust to a new equilibrium
state where analytical solutions are not available.
Right afterwards, we apply an external poloidal field using a vector
potential generalized from \citet{Zanni_etal07}
\begin{equation}
A_\phi(r,\theta)=\frac{2B_{z0}r_{\rm in}}{4-\alpha-q_T}
\bigg(\frac{R}{r_{\rm in}}\bigg)^{-\frac{\alpha+q_T}{2}+1}
[1+(m\tan\theta)^{-2}]^{-\frac{5}{8}}\ ,\label{eq:Aphi}
\end{equation}
where $m$ is a parameter that controls how much poloidal fields are bent, with
$m\rightarrow\infty$ giving a pure vertical field. We choose $m=1$, and we have
verified that the results are insensitive to the choice of $m$.
Poloidal fields are obtained from ${\mb B}=\nabla\times(A_{\phi}\hat{\phi})$ in a
way that guarantees $\nabla\cdot{\mb B}=0$. At the midplane, we have
${\mb B}=B_{z0}\hat{z}(r/R_0)^{-(\alpha+q_T)/2}$, maintaining constant ratio of
gas to magnetic pressure, defined by plasma $\beta_0$.
Fiducially, we choose $\beta_0=10^5$, appropriate for the inner
region of PPDs \citep{BaiStone13b,Bai13}, but we also consider stronger
fields in Section \ref{ssec:beta}.

In simulations including the Hall effect with aligned field polarity, we find that the
outcome of the simulation depends on initial conditions. This issue is discussed
in more detail in Appendix \ref{app:init}. For simulations shown in the main text,
we have modified the simulation setup to obtain more consistent result (see
Section \ref{sec:align} for details).

An important aspect of the simulations is to properly control the inner boundary
condition. We note that the flow near the polar region should originate from a
part of the disk located inside the inner boundary, whose dynamics is beyond the
reach of the simulation. A standard outflow-type boundary condition would
violate causality, and can become unstable in the presence of magnetic fields
which further interfere with the wind flow in the main computational domain. After
experimenting with a number of options, we adopt a boundary condition that is
close to a fixed state, which we find can better constrain the flow structure
near the inner boundary and minimize its effect to the rest of the
simulation domain.  We fix the density profile as (\ref{eq:rho}) with a density floor,
$v_r=v_\theta=0$, and $v_\phi$ is set to be the minimum of the initial $v_\phi$
(\ref{eq:vphi}) and $\Omega_K(r_{\rm in})R$. Gas temperature is set based on
(\ref{eq:fullepsi}). We also set a buffer zone between $r=r_{\rm in}$ and
$r=1.5r_{\rm in}$, where we linearly reduce all magnetic diffusivities with radius to
zero, and damp gas poloidal velocities on local orbital timescale. This approach
prevents accretion into the inner boundary and would lead to some mass
accumulation within the buffer zone. We then deplete the gas in the buffer zone
over a timescale of $10^4\Omega_K^{-1}(r_{\rm in})$. Also note that the duration
of our simulations is relatively short, and we do not observe significant
modifications of the inner disk structure.

The rest of the boundary conditions are straightforward.
The outer radial boundary follows from standard outflow boundary prescriptions, where
hydrodynamic variables are copied from the last grid zone assuming
$\rho\propto r^{-2}$, $v_\phi\propto r^{-1/2}$, with $v_r$ and $v_\theta$ unchanged
except that we set $v_r=0$ in case of inflow.
Magnetic variables in the inner/outer ghost zones are copied from the nearest grid zone
assuming $B_r\propto r^{-2}$ and $B_\phi\propto r^{-1}$, with $B_\theta$ unchanged.
Reflection boundary conditions are applied in the $\theta-$boundaries.

\subsection[]{List of Runs and Parameters}

\begin{table}
\caption{List of Simulation Runs}\label{tab:runlist}
\begin{center}
\begin{tabular}{c|c|cc|c}\hline\hline
 Run & Polarity &  $\beta_0$ & $\Sigma_{\rm FUV}$ & Runtime ($\Omega_0^{-1}$)\\\hline
Fid+ & + & $10^5$ & $0.03$ & 22500 \\%& $10^{-4}$ \\
Fid0 & No Hall & $10^5$ & $0.03$ & 15000 \\% & $10^{-4}$ \\
Fid$-$ & $-$ & $10^5$ & $0.03$ & 24000 \\\hline% & $10^{-4}$ \\\hline
B4+ & + & $10^4$ & $0.03$ & 9600 \\% & $10^{-4}$ \\
B40 & No Hall & $10^4$ & $0.03$ & 6000 \\% & $10^{-4}$ \\
B4$-$ & $-$ & $10^4$ & $0.03$ & 15000 \\\hline% & $10^{-4}$  \\\hline
FUV+ & + & $10^5$ & $0.3$ & 8100 \\% & $10^{-4}$ \\
FUV0 & No Hall & $10^5$ & $0.3$ & 12000 \\% & $10^{-4}$ \\
FUV$-$ & $-$ & $10^5$ & $0.3$ & 6000 \\\hline% & $10^{-4}$ \\\hline
%NoGr+ & + & $10^5$ & $0.03$ & No \\
%NoGr0 & No Hall & $10^5$ & $0.03$ & No \\
%NoGr$-$ & $-$ & $10^5$ & $0.03$ & No \\\hline
\end{tabular}
\end{center}
All other parameters are fixed, with $M_*=M_\odot$,
$\Sigma=500R_{\rm AU}^{-1}$g cm$^{-2}$, $H_d/R=0.045R_{\rm AU}^{1/4}$.
The simulation domain extends from 0.6-60 AU in radius.
\end{table}

We list all our simulation runs in Table \ref{tab:runlist}. Most parameters are fixed as
described in previous subsections, and we only vary two parameters: disk
magnetization (parameterized by plasma $\beta_0$, fiducially $10^5$), and FUV
penetration depth $\Sigma_{\rm FUV}$ (fiducially $0.03$g cm$^{-2}$).
Time is measured in unites of $\Omega_0^{-1}\equiv\Omega_K(r_{\rm in})^{-1}$
in our simulations. For $r_{\rm in}=0.6$ AU, we have $\Omega_0^{-1}=0.074$ yr.
We will focus on our fiducial runs, labeled as ``Fid$\pm$",
where the $+$/$-$ signs correspond to simulations with poloidal field
aligned/anti-aligned with disk rotation. For comparison, we also conduct
a run ``Fid0", where we turn off the Hall effect. These simulations are run for more
than 2000 orbits at innermost radius for detailed analysis.
We then vary one parameter at a time, and for each variation, three runs labeled by
``$\pm0$" are performed as in the fiducial case. They are run for shorter amount of
time but are sufficient to illustrate the dominant features. 

\section[]{Diagnostics}\label{sec:diag}

In this section, we discuss major diagnostics to be employed to analyze our
simulation results.

\subsection[]{Elsasser Numbers}

The strength of the non-ideal MHD effects are conveniently measured by
dimensionless Elsasser numbers., defined as
\begin{equation}
\Lambda\equiv\frac{v_A^2}{\eta_O\Omega_K}\ ,\quad
\chi\equiv\frac{v_A^2}{\eta_H\Omega_K}\ ,\quad
Am\equiv\frac{v_A^2}{\eta_A\Omega_K}\ ,
\end{equation}
where $v_A\equiv B/\sqrt{4\pi\rho}$ is the Alfv\'en speed.
Non-ideal MHD terms are considered strong if the Elsasser numbers are
of order unity of less. For Ohmic resistivity, $\Lambda<1$ is generally
sufficient to suppress the MRI \citep{Turner_etal07,IlgnerNelson08}. For
AD, $Am<1$ can suppress or damp the MRI depending on vertical field
strength \citep{BaiStone11}.

Since $\eta_O$ is constant, $\eta_H\propto B$ and $\eta_A\propto B^2$,
only $Am$ is independent of field strength. For the Hall effect, a field
strength independent measure is the Hall length $l_H$, defined as
\begin{equation}
l_H\equiv \eta_H/v_A\ ,
\end{equation}
which is the generalization of ion inertial
length in weakly ionized plasmas \citep{KunzLesur13}. Strong Hall
effect is characterized by $l_H\gtrsim H$.

\subsection[]{Angular Momentum Transport and Disk Flow Structure}

Angular momentum transport is mainly mediated by magnetic stresses.
Ignoring hydrodynamic processes, the equation of
angular momentum transport in cylindrical coordinates can be written as
\begin{equation}\label{eq:accrate}
\frac{\dot{M}_{\rm acc}v_K}{4\pi}
=\frac{\pa}{\pa R}\bigg(R^2\int_{-z_b}^{z_b} dz\overline{T_{R\phi}}\bigg)
+R^2\overline{T_{z\phi}}\bigg|^{-z_b}_{z_b}\ ,
\end{equation}
where $\dot{M}_{\rm acc}\equiv-2\pi R\int_{-z_b}^{z_b}\rho v_Rdz$ is the accretion rate,
$T_{R\phi}\equiv -B_RB_\phi/4\pi$, $T_{z\phi}\equiv-B_zB_\phi/4\pi$ are
Maxwell stresses, and $\pm z_b$ mark the vertical coordinates that separate
the disk and atmosphere. Overlines represent time and azimuthal
averages, and we have assumed Keplerian rotation.

Physically, the first term on the right hand side corresponds to
radial transport of angular momentum. In our simulations, the disks
are largely laminar, and this term is dominated by large-scale fields
that wind up into spirals, corresponding to magnetic braking.
By convention, we define the \citet{ShakuraSunyaev73} $\alpha$
parameter as
\begin{equation}\label{eq:alpha}
\alpha\equiv\int_{-z_b}^{z_b}T_{R\phi}dz/\int_{-z_b}^{z_b}Pdz\ ,
\end{equation}
which is dimensionless measure of the stress.
Note that the accretion rate is related to the radial gradient of
$T_{R\phi}$.

The second term on the right hand side corresponds to vertical transport
of angular momentum by magnetized disk winds. We can normalize
$T_{z\phi}$ by the midplane gas pressure. We note that given
similar field strength, vertical transport is more efficient than radial
transport by a factor of $\sim R/H$ \citep{Wardle07,BaiGoodman09}.
The $B_zB_\phi$ stress drives accretion by exerting a torque on the
disk, and the torque density is proportional to its vertical gradient.
Because $B_z\approx$ constant in a thin disk, the wind-driven local
accretion velocity is given by \citep{BaiStone13b}
\begin{equation}\label{eq:radflow}
-\frac{1}{2}\rho\Omega_Kv_R\approx-\frac{B_z}{4\pi}\frac{dB_\phi}{dz}\ .
\end{equation}
We see that the accretion mass flux $\rho v_R$ is directly proportional to
toroidal field gradient. This is the most important relation for understanding
the flow structure in our simulations.

\begin{figure*}
    \centering
    \includegraphics[width=180mm]{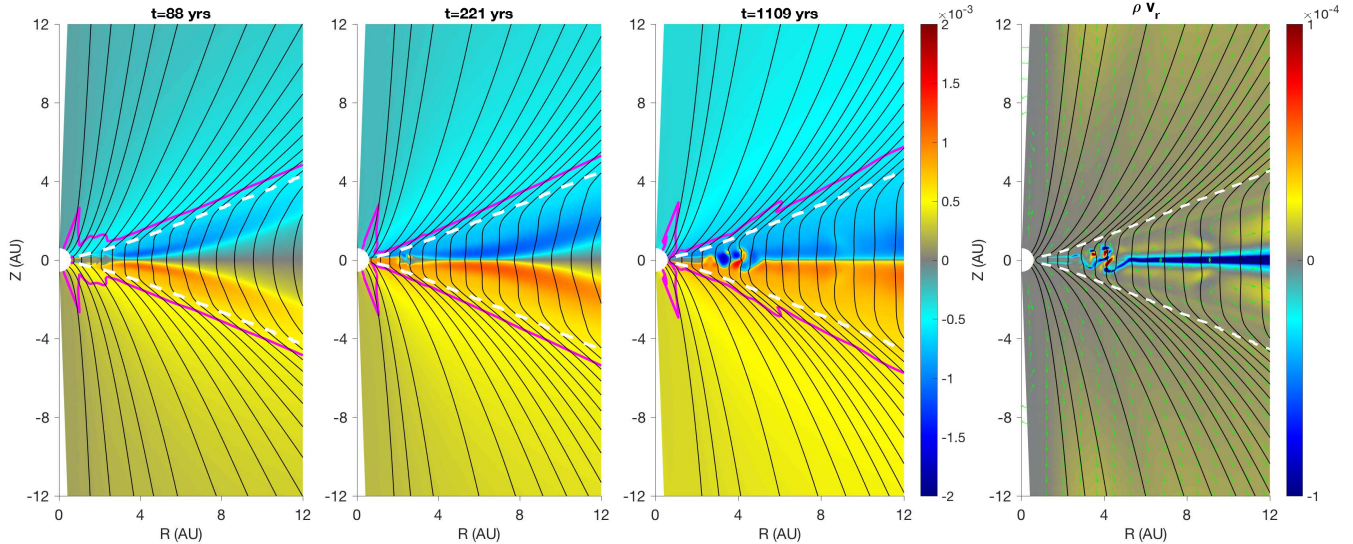}
  \caption{Result from the Hall-free simulation Fid0.
  Left three panels: snapshots of magnetic field configuration represented
  by equally-spaced contours of poloidal magnetic flux and color (scaled toroidal
  field $RB_\phi$) at $t=$1200, 3000 and 15000 $\Omega_0^{-1}$. Rightmost panel:
  radial mass flux $\rho v_r$ (rescaled by $r^{q_D}R^{1/2}$ at the last snapshot
  (t=15000$\Omega_0^{-1}=1109$ yrs), overlaid with poloidal flux contours (black)
  and velocity vectors (green arrows). Thick white dashed lines in the four panels mark
  the FUV ionization front, and the magneta contours in the left three panels mark the
  Alfv\'en surface. Note that the simulation domain extends to $r=60$ AU.}\label{fig:nh_bevolve}
\end{figure*}

\subsection[]{Wind Kinematics}\label{ssec:windkin}

In steady state and axisymmetry, a magnetized disk wind is characterized
by a series of conservation laws along field lines
(e.g., \citealp{Spruit96}), as long as the gas is
well coupled with magnetic field (i.e., ideal MHD). These include the
conservation of mass 
\begin{equation}
k\equiv\frac{4\pi\rho v_p}{B_p}\ ,
\end{equation}
angular velocity of magnetic flux surface
\begin{equation}
\omega\equiv\Omega-\frac{kB_\phi}{4\pi\rho R}\ ,
\end{equation}
and specific angular momentum
\begin{equation}
l\equiv\Omega R^2-\frac{RB_\phi}{k}\ ,
\end{equation}
subscript $_p$ denotes the poloidal component.
Here, $k$, $\omega$ and $l$ are conserved along poloidal field lines,
and $\Omega\equiv v_\phi/R$. In practice, we normalize these quantities to
$k_0=4\pi\rho_{\rm mid}v_K/B_{z0}$, $\Omega_K$, and $\Omega_KR_0^2$,
where $v_K$, $\Omega_K$ and $B_{z0}$ are defined at the wind launching
radius $R_0$. If the equation of state is barotropic (ours is not), energy
conservation can also be expressed explicitly.
We will test the above three relations in our simulations.

The most important wind diagnostics is the mass loss rate. We define
$\dot{M}_{\rm wind}(R)$ as the cumulative wind mass loss rate within
radius $R$. Locally, we quote the mass loss rate per logarithmic radius as
\begin{equation}\label{eq:mloss}
\frac{d\dot{M}_{\rm wind}}{d\ln R}
=2\pi R^2[(\overline{\rho v_z})_{z_b}+(-\overline{\rho v_z})_{-z_b}]\ .
\end{equation}

An important concept is the Alfv\'en radius $R_A$, where for the wind
flow originating from radius $R_0$, $R_A$ is the radius of the point along
the field line where the poloidal flow velocity $v_p$ equals to the poloidal
Alfv\'en velocity $v_{Ap}=B_p/\sqrt{4\pi\rho}$.
The local wind mass loss rate is closely related to wind-driven accretion
rate by \citep{FerreiraPelletier95,Bai_etal16}
\begin{equation}\label{eq:lever}
\frac{d\dot{M}_{\rm wind}}{d\ln R}
=\frac{\dot{M}_{\rm acc}}{2}\frac{1}{(R_A/R_0)^2-1}\ .
\end{equation}
Therefore, the location of the Alfv\'en point provides an alternative and
more convenient measure of the wind mass loss rate. The ratio
$\lambda\equiv(R_A/R_0)^2$ is defined as the magnetic lever arm.

\subsection[]{Magnetic Flux Transport}

The basic physics of magnetic flux transport due to non-ideal MHD
effects has been studied in detail in \citet{BaiStone17}, and we do not pursue
analysis in as much detail as was done there. In most occasions, we simply
measure the total magnetic flux enclosed within radius $r$ at the midplane
\begin{equation}\label{eq:PhiB}
\Phi_{B,{\rm mid}}(r)=2\pi\int_{0}^{\pi/2}B_r(r,\theta)r^2\sin\theta d\theta\ ,
\end{equation}
and follow its evolution.

\section[]{Benchmark: The Fiducial Hall-free Simulation}\label{sec:nohall}

We start from the Hall-free simulation before introducing further complications
owing to the Hall effect.  
In Figure \ref{fig:nh_bevolve}, we show snapshots of magnetic field configurations
from run Fid0. Overall, the system quickly settles into a
laminar configuration in approximately steady state in 10-15 local orbital time,
launching a disk wind. To a large extent, the system is symmetric about the midplane
(except for regions between $R\sim2-5$ AU). We will further discuss the field
configuration in Section \ref{ssec:nh_field}. 

The quasi-steady state configuration allows us to choose two characteristic radii,
2 AU and 10 AU, and analyze the overall gas dynamics in further detail.
In Figures \ref{fig:elsprof} and \ref{fig:nh_prof},
we show the vertical profiles of main diagnostic quantities at the
last simulation snapshot ($t=15000\Omega_0^{-1}\approx$1109 yrs), including
density, temperature, Elsasser numbers, pressure, Maxwell stress, magnetic fields
and velocity fields. The results are discussed in Sections \ref{ssec:elsasser},
\ref{ssec:nh_field} and \ref{ssec:nh_flow} from different perspectives. Section \ref{ssec:nh_flow} further
addresses the overall disk angular momentum transport and examines the disk
flow structure. We analyze wind kinematics in Section \ref{ssec:wind}.

\subsection[]{Magnetic Diffusivities}\label{ssec:elsasser}

\begin{figure}
    \centering
    \includegraphics[width=90mm]{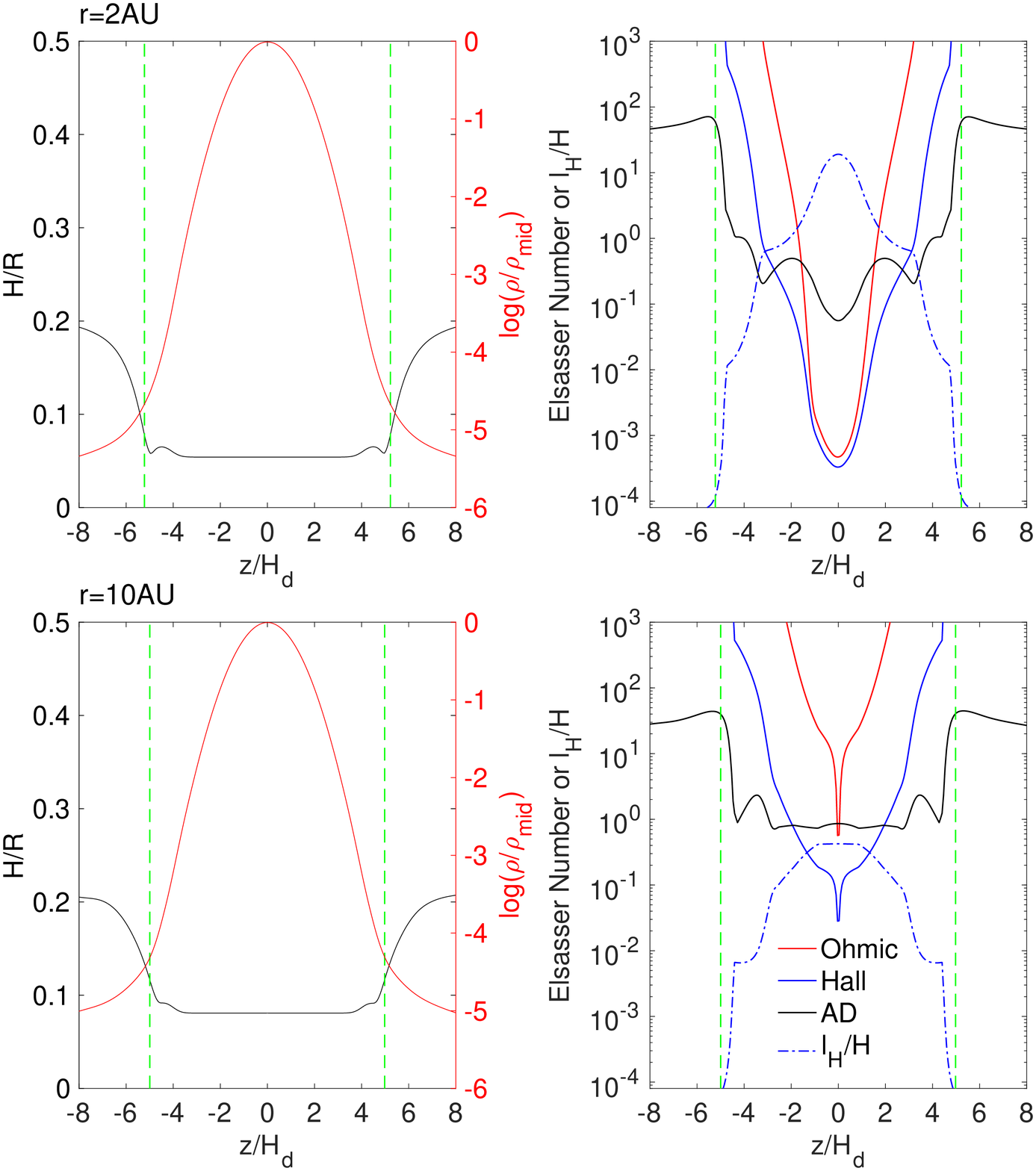}
  \caption{Vertical profiles of basic diagnostic quantities from the Hall-free
  simulation Fid0 at cylindrical radii $R=2$ AU (upper panels) and
  $R=10$ AU (lower panels), measured at the last snapshot of the simulation
  (t=15000$\Omega_0^{-1}\approx$1109 yrs). Left panels show the density (red) and temperature
  (black, expressed in $H/R$) profiles. Right panels show the three non-ideal MHD
  Elsasser numbers (solid) and the normalized Hall length $l_H/H$ (dash-dotted). The
  vertical green dashed lines mark the location of the FUV ionization front.}\label{fig:elsprof}
\end{figure}

\begin{figure*}
    \centering
    \includegraphics[width=180mm]{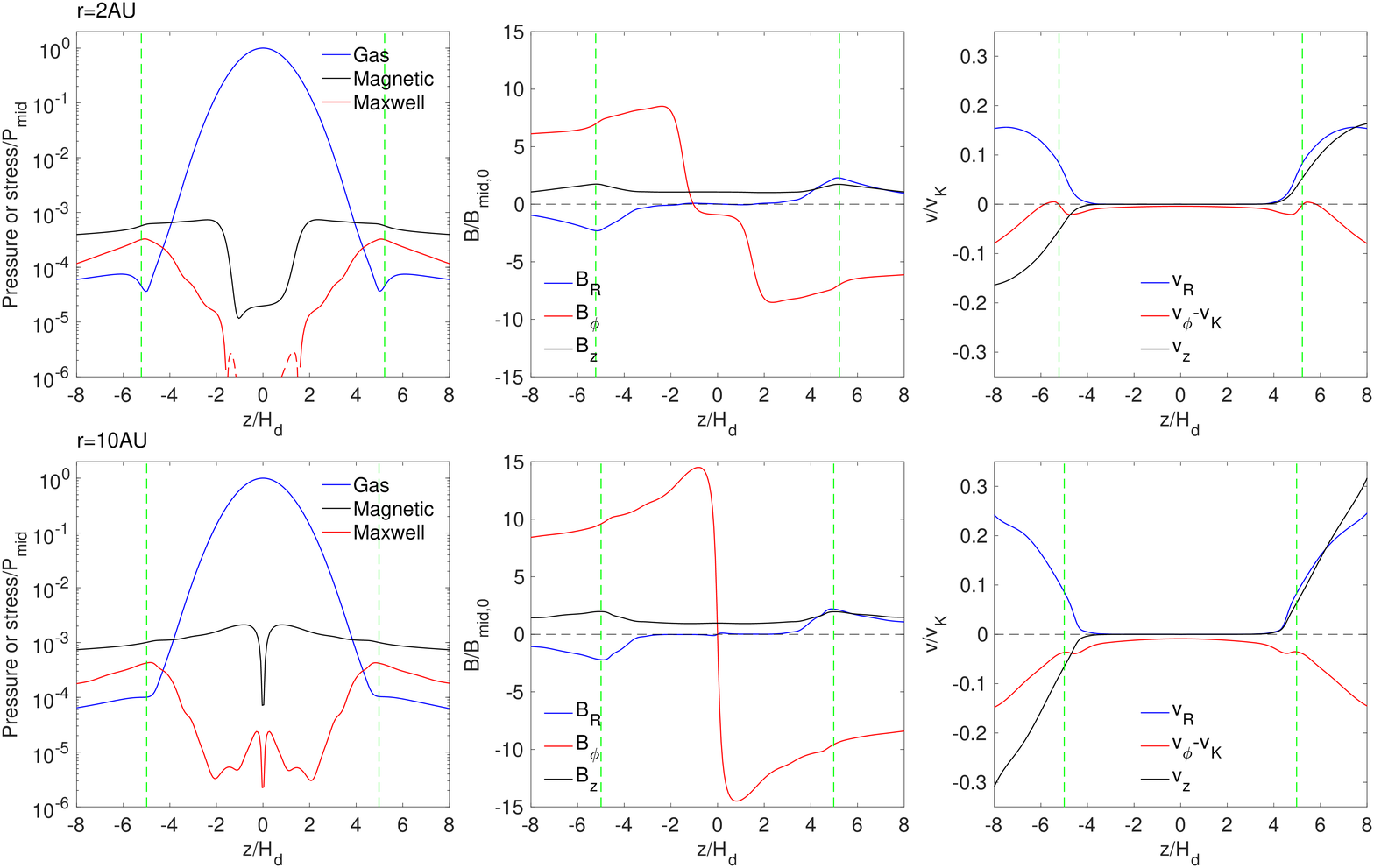}
  \caption{Vertical profiles of main diagnostic quantities from the Hall-free
  simulation Fid0 at cylindrical radii $R=2$ AU (upper panels) and
  $R=10$ AU (lower panels), measured at the last simulation snapshot
  ($t=15000\Omega_0^{-1}\approx$1109 yrs). Left panels show the gas pressure (blue),
  magnetic pressure (black) profiles, together with the profile of Maxwell stress
  $T_{R\phi}=-B_RB_\phi/4\pi$. Middle panels show the profiles of the three components of
  the magnetic field. Right panels show the three components of gas velocity (Keplerian
  rotation subtracted), with the inset zooming in to highlight the accretion flow. The
  vertical green dashed lines mark the location of the FUV ionization front.}\label{fig:nh_prof}
\end{figure*}

From Figure \ref{fig:elsprof}, we see that at both $R=2$ and $10$ AU, the FUV front
is located around $z=\pm5H_d$.
Our thermodynamic scheme nicely maintains constant disk aspect ratio
$\epsilon=\epsilon_d(R)$ within the disk zone, and allows it to smoothly rise to
$\epsilon\sim0.2$ beyond the FUV front. The density decreases with height much
more slowly near and beyond the
the FUV front, owing to magnetic pressure support (see Figure \ref{fig:nh_prof}).

The right panels of Figure \ref{fig:elsprof} show the Elsasser number profiles.
Note that while we have switched off the Hall effect, we can still show the Hall
Elsasser numbers and the Hall length $l_H$. Also, $l_H$ and the AD Elsasser
number $Am$ profiles, being independent of field strength, remain largely
unchanged over the course of the simulation. This also holds in simulations with
the Hall term turned on (we thus do not repeat similar plots in later discussions).

We first focus on the ambipolar Elsasser number $Am$.
The FUV front leads to the most sharp increase of $Am$ to near $\sim100$,
making ideal MHD a good approximation in the wind zone. Below the FUV front,
the $Am$ profiles display a number of wiggles, corresponding to contributions from
individual ionization sources (direct X-ray absorption, X-ray scattering, cosmic-rays).

At 2 AU, resistivity and the Hall effect are the two dominant non-ideal MHD effects
at the midplane, with Elsasser number orders of magnitudes below $1$. This
corresponds to the conventional Ohmic dead zone \citep{Gammie96}, with extremely
weak level of ionization with magnetic field largely decoupled with the gas.
The Hall effect dominates between $z\sim1-3H_d$, and AD takes over to dominate at
disk upper layers. At 10 AU, Ohmic resistivity becomes largely irrelevant in the entire
vertical column. The Hall effect dominates the midplane region, and AD dominates
beyond $z\sim\pm2H_d$.  Note that Hall length $l_H$ well exceeds $H_d$ at $R=2$ AU
in the midplane region, whereas it drops below $H_d$ at $10$ AU, reflecting that the
Hall effect weakens towards larger disk radii.

The Elsasser number profiles obtained here are largely consistent with those obtained
in local simulations for the inner disk using similar parameters \citep{BaiStone13b,Bai13}.
We also note a flat $Am$ profile $\sim1$ at $10$ AU within the FUV front. This fact also
approximately holds towards larger radii.

\subsection[]{Magnetic Field Configuration}\label{ssec:nh_field}

The vertical structure of magnetic fields in the disk can be seen from Figure
\ref{fig:nh_prof}. Overall, the field configuration is consistent with previous
(vertically-)local simulations
(e.g., Figure 11 of \citealp{BaiStone13b}, Figure 6 of \citealp{Gressel_etal15}).

Initially, the poloidal fields bend radially outwards,\footnote{In the Hall-free simulations,
the final field configuration is insensitive to initial conditions, though having an outward
bent initial configuration is the most natural.}
generating oppositely-directed toroidal fields above/below the midplane, which build
up magnetic pressure. As seen from the Figure, magnetic pressure dominates over
gas pressure beyond about $z=\pm4H_d$. The vertical gradient of toroidal field
exerts a torque to the disk, driving radial flows according to (\ref{eq:radflow}). The
radial flow is an accretion flow within a certain height about the disk midplane
before $|B_\phi|$ reaches its maxima. Beyond the maxima, $|B_\phi|$ slowly
decreases with height (driving a weak decretion flow), and its associated magnetic
pressure gradient directly drives the disk wind (see Section \ref{ssec:wind}).

Toroidal field is always the dominant field component.
In steady state, the generation of toroidal field (from radial field) is mainly balanced
by Ohmic/ambipolar dissipation in the bulk disk, whereas in the wind zone, it is
simply balanced by advection in the wind flow. The poloidal field configuration,
especially the level field lines bend, is mainly set by the radial flow structure through
the vertical extent of the disk (which transitions from accretion to decretion towards
surface, and drives field line bending), Ohmic/ambipolar dissipation
(which works to straighten the field in the bulk disk), as well as outflow advection
(towards the surface).

\begin{figure}
    \centering
     \subfigure{
    \includegraphics[width=80mm]{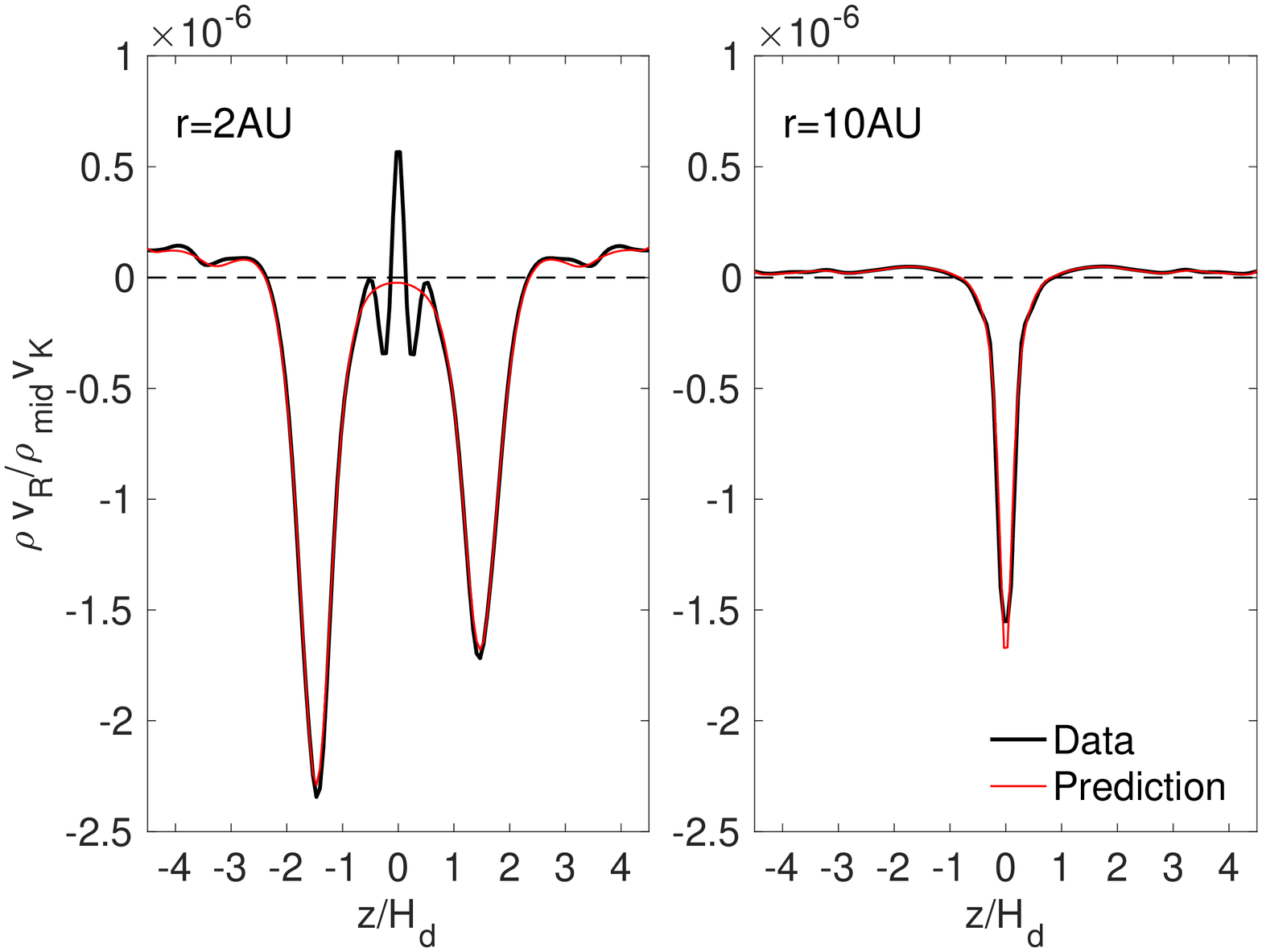}}
    \subfigure{
    \includegraphics[width=90mm]{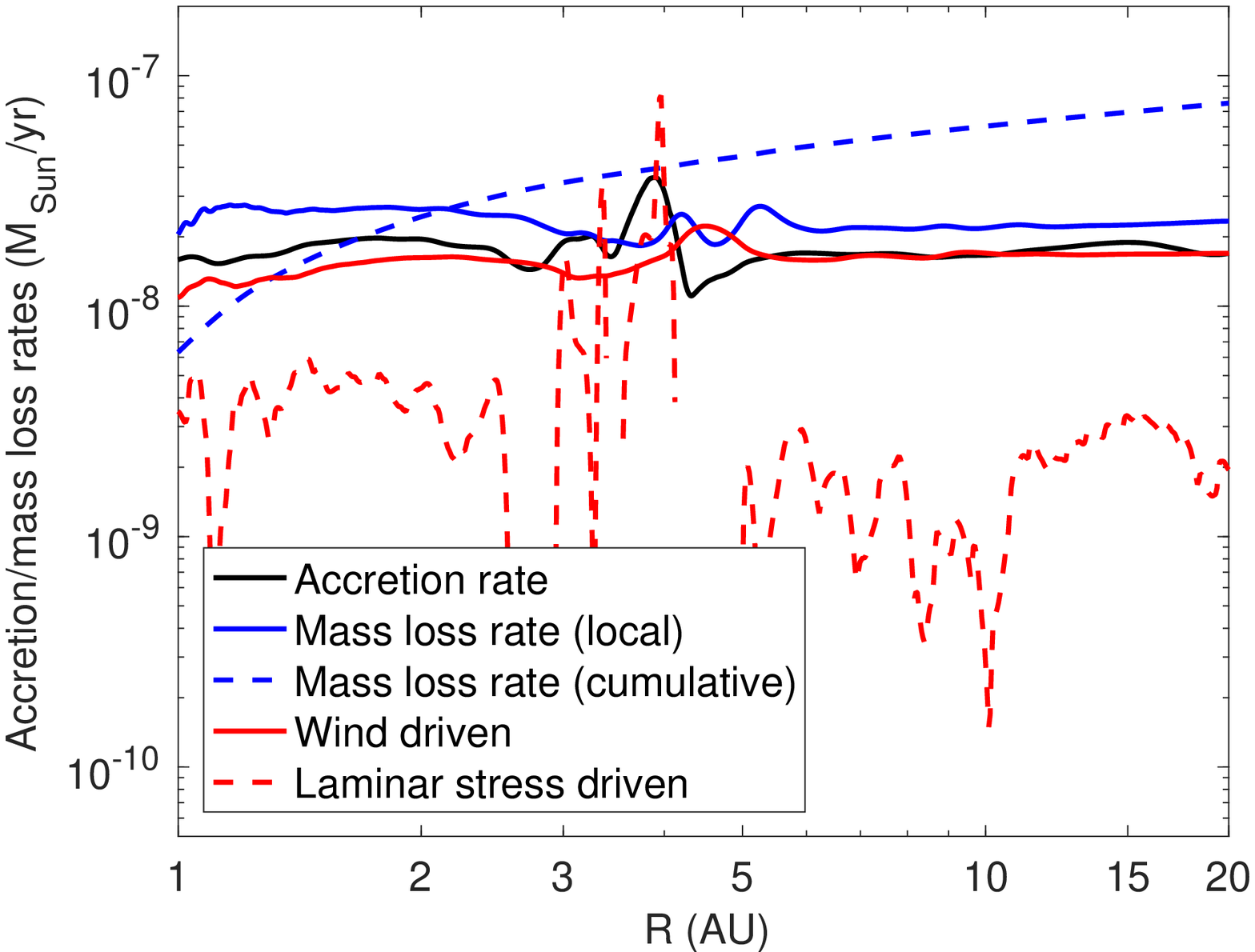}}
  \caption{Top panels: vertical distribution of radial mass flux measured at two radii $R=2$
  AU (left) and $10$ AU (right) at the end of the Hall-free simulation Fid0 (black), together with
  predicted mass flux based on Equation (\ref{eq:radflow}). The horizontal dashed line marks
  zero mass flux to guide the eye. Bottom panel: radial profile of
  mass accretion rate (black), mass loss rate per logarithmic radius (blue solid), cumulative
  mass loss rate (blue dashed), predicted wind-driven accretion rate (red solid), and the
  predicted viscously-driven accretion rate (red dashed). All calculated from the last snapshot
  of the simulation.}\label{fig:nh_acc}
\end{figure}

\subsubsection[]{The Symmetry Issue}\label{sssec:symmetry}

Launching MHD wind with physical symmetry requires toroidal field to change
sign across the disk, which is directly connected to driving the accretion flow
\citep{BaiStone13b}. This is achieved in different ways at small and
large radii.

At small radii (e.g., $R\lesssim2$ AU), the extremely weak
level of ionization means that there are very few charge carriers to sustain current, which
minimizes the vertical gradient of toroidal field. In \citet{BaiStone13b}, we found that
toroidal field of one sign reaches maxima has a flat profile across the midplane, it then
flips sharply at a few scale heights at the other side the midplane (as the gas becomes
better coupled to the field). This is observed in early stages of evolution (i.e., second
panel in Figure \ref{fig:nh_bevolve}). Later, on the other hand, midplane toroidal
field decreases, and toroidal field peaks at a similar heights {\it both}
above {\it and} below the midplane (i.e., third panel in Figure \ref{fig:nh_bevolve},
and central top panel of Figure \ref{fig:nh_prof}). These are similar to the ``belt" structures
observed in \citet{Gressel_etal15}.

At large radii (e.g., $R\gtrsim5$ AU), the midplane region becomes better
coupled with magnetic field (Elsasser number $\sim1$), allowing the flip of toroidal field to
take place right at
the midplane. This leads to reflection symmetry about the midplane, as seen in Figure
\ref{fig:nh_bevolve}, and the central bottom panel of Figure \ref{fig:nh_prof}. This result
is consistent to earlier local studies of \citet{Bai13}.

In between $R\sim2-5$ AU, the midplane region of the disk is dominated by patches
of toroidal field in opposite signs. These patches interact with each other, leading
to some secular evolution of the system. We do not find signatures of unstable MRI
channel modes develop, as in some runs in \citet{Gressel_etal15}. Instead, we simply
interpret these phenomena as inherent to the transition from the extremely poorly coupled
regime at small radii to the marginally coupled regime at larger radii.\footnote{On the other
hand, upon running this simulation further, the MRI starts to slowly develop at
$R\gtrsim9$ AU, leading to further distortions of fields and flow structure, which would
require 3D simulations to capture properly.}

\subsection[]{Angular Momentum Transport and Flow Structure}\label{ssec:nh_flow}

Magnetized disk wind is the dominant mechanism of disk angular momentum transport.
The vertical distribution of wind-driven accretion flow directly results from toroidal field
gradient, shown in Equation (\ref{eq:radflow}). This is verified in the top panels of Figure
\ref{fig:nh_acc} for $R=2$ and $10$ AU, respectively. At $R=2$ AU, there are mainly
two accretion layers above and below the midplane at around $z=\pm1.5H_d$, where
maximum toroidal field gradient develops, with slightly different mass fluxes (owing to
slight asymmetry in $B_\phi$, as seen in the central top panel of Figure \ref{fig:nh_prof}).
At $R=10$ AU, the accretion flow is confined at the midplane, where toroidal field flips
symmetrically.

The radial flow velocity can be found in the insets on the right panels of Figure
\ref{fig:nh_prof}. The thickness of the accreting layer is typically a good fraction of
a scale height, and is well resolved in the simulation. We note that beyond the
accreting layer towards the surface, the gas is directed radially outward (i.e., decretion)
because the toroidal field gradient reverses. Nevertheless, combined with the rapid
density drop, the surface layer decretion carries a negligible fraction of mass flux,
as seen in Figure \ref{fig:nh_acc}.

The transition to ideal MHD regime at the FUV front shows distinct features in
magnetic field and flow structures, as can be found in Figure \ref{fig:nh_prof},
and they are mostly consistent with previous local simulations
\citep{BaiStone13b,Bai13}. In particular, we have defined the wind base as the
location where the gas flow transitions from being sub-Keplerian (below) to
super-Keplerian (above)
\citep{WardleKoenigl93}, and the location is found to largely coincide with
the FUV front \citep{BaiStone13b,Gressel_etal15}.
However, in our global simulations, we find that such transition never takes place.
Instead, $v_\phi$ almost never exceeds Keplerian, and simply decreases towards
larger height. This is related to the nature of magneto-thermal disk winds
\citet{Bai_etal16}, to be further discussed in Section \ref{ssec:wind}.
On the other hand, the FUV front does correspond to a local maxima in $v_\phi$,
which may still be considered as a reasonable way to define the wind base $z_b$,
as we adopt here.
Similar situation holds for simulations with the Hall effect\footnote{Except that when
the strong current layer is located close to the FUV front, $v_\phi$ there can be
strongly reduced.}.

\begin{figure*}
    \centering
    \includegraphics[width=160mm]{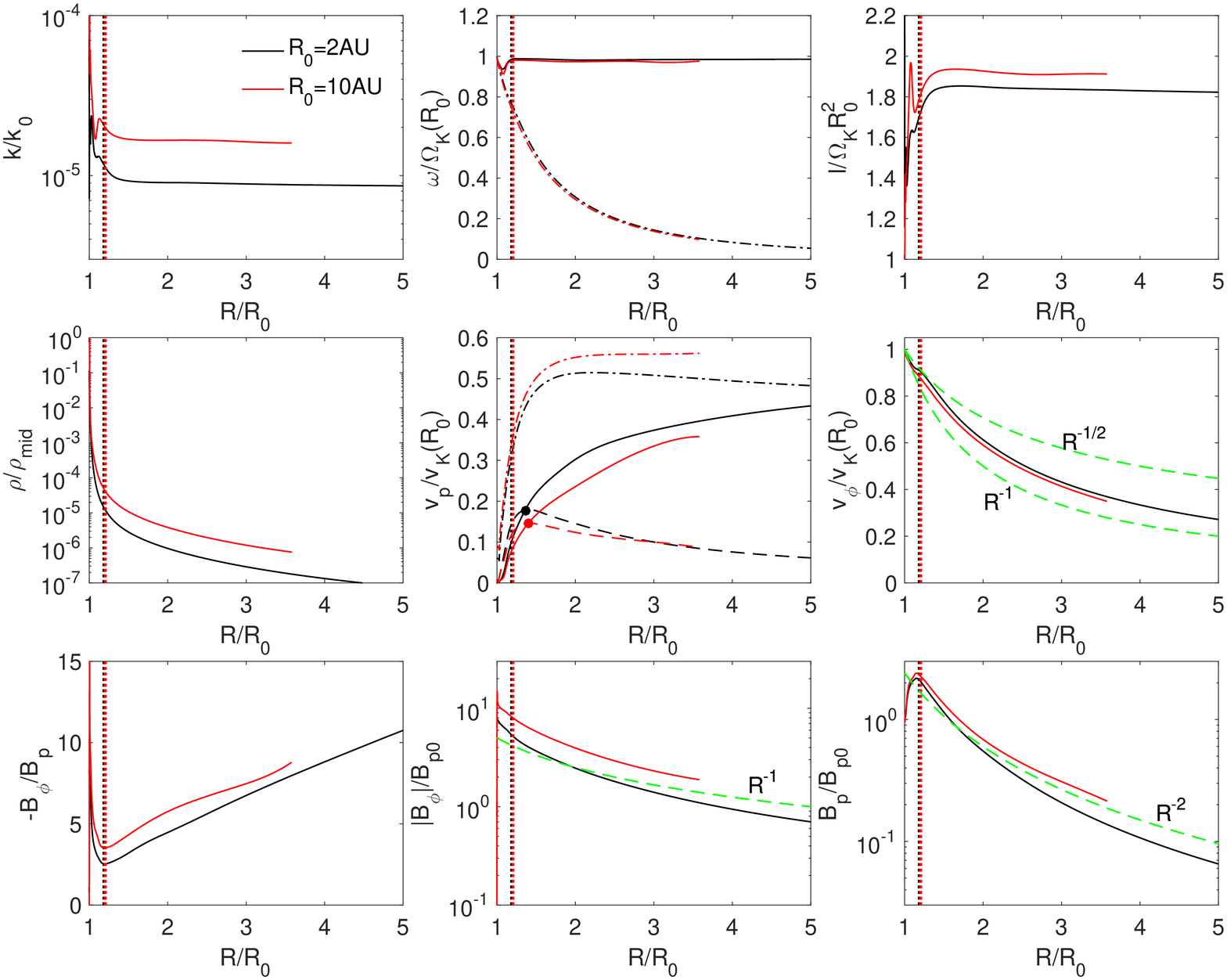}
  \caption{Wind properties in the Hall-free simulation Fid0, traced along two
  representative field lines originating from $R_0=2$ AU (black) and $10$ AU (red),
  respectively.
  Top three panels show the three conserved quantities $k$, $\omega$, and $l$,
  normalized to local values, outlined in Section \ref{ssec:windkin}. The middle
  three panels show the gas density $\rho$, poloidal velocity $v_p$ and toroidal
  velocity $v_\phi$ along the wind field lines. In the central panel, we further plot the
  poloidal Alfv\'en speed $v_{Ap}$ and poloidal fast magnetosonic speed $v_{fp}$ in
  dashed and dash-dotted lines. Filled circles mark the locations of the
  Alfv\'en points. The bottom three panels show the ratio of
  toroidal-to-poloidal field $|B_\phi|/B_p$, as well as toroidal and poloidal fields
  normalized to the poloidal field strength at the midplane. Vertical dashed lines mark
  the location of the FUV front along the each of the field lines. Red dots on the left
  panels mark the locations of the Alfv\'en points.}\label{fig:windprof}
\end{figure*}

The bottom panel of Figure \ref{fig:nh_acc} shows the radial profiles of
accretion and mass loss rates, together with accretion rates computed
from (\ref{eq:accrate}), separating the contribution from wind-driven
transport from radial transport of angular momentum. Overall, the mass
accretion rate is approximately constant over radius, and is around
$2\times10^{-8}M_{\bigodot}$ yr$^{-1}$. Note that over the
course of our simulation, there has been very little evolution of gas surface
density. The flat accretion rate profile is largely a result of proper choice
of the initial magnetic flux distribution (i.e., constant plasma $\beta_0$).

We again see that disk wind accounts for almost the entire accretion process
in the disk, whereas radial transport of angular momentum due to $T_{R\phi}$
from the Maxwell stress is about one order of magnitude smaller. The
Shakura-Sunyaev $\alpha$ measured from our simulation is small, ranging
from $\alpha\sim10^{-4}$ at $R\sim1$ AU, to $\alpha\sim4\times10^{-4}$ beyond
$R=5$ AU. Such small value can already be inferred from the left panels of
Figure \ref{fig:nh_prof}.

At intermediate radii between $\sim2-5$ AU, the
system exhibits slightly larger accretion rates, which is related to enhanced
magnetic activities associated with the secular evolution of toroidal field
patches. We also see from the rightmost panel of Figure \ref{fig:nh_bevolve}
that the disk gas shows complex radial flow structures. In these regions, 
(\ref{eq:accrate}) is no longer applicable to predict accretion rates due to
time variability, but overall, the accretion flow still largely correlates with
strong toroidal field gradients at the boundaries of the toroidal field patches,
where most of the torque is exerted.

\subsection[]{Wind Kinematics}\label{ssec:wind}

Figure \ref{fig:nh_acc} further demonstrates that wind mass loss rate, calculated
from (\ref{eq:mloss}), is excessive.
The mass loss rate per logarithmic radii already exceeds the accretion rate,
whereas the cumulative mass loss rate reaches as large as 4 times the
accretion rate at $R=20$ AU. This is again a consequence of magneto-thermal
disk winds, which we focus on in this subsection.

To analyze the wind kinematics, we again choose $R_0=2$ and $10$ AU, and trace
poloidal field lines from the midplane all the way to the boundary of our simulation
domain. We measure various diagnostic quantities along the field lines and show
the results in Figure \ref{fig:windprof}.

We see that the conservation laws outlined in Section \ref{ssec:windkin} are generally
well satisfied beyond the FUV front where the gas behaves approximately in the ideal
MHD regime. The mass flux along field lines is substantial. With
$k\sim10^{-5}k_0$ and $H/R\sim0.05-0.08$ (at $R=2-10$ AU), the corresponding local
disk depletion timescale is typically only a few thousand orbits.
We also find that $\omega\approx\Omega_K(R_0)$ as the angular velocity of magnetic
flux surface. The value of $l\lesssim2\Omega_K(R_0)R_0^2$, indicating that the wind
carries less than twice the specific angular momentum at the origin, which is consistent
with the fact that local wind mass loss rate is comparable to wind-driven accretion rate.

Poloidal velocity constantly increases along the field line within the simulation domain.
We see in Figure \ref{fig:nh_bevolve} that the Alfv\'en surface is located relatively close
to the disk, and is only slightly beyond the FUV front. The fast magnetosonic point,
defined when poloidal velocity equals to the poloidal fast magnetosonic velocity
$v_{fp}^2=(1/2)[(c_s^2+v_A^2)+\sqrt{(c_s^2+v_A^2)^2-4c_s^2v_{Ap}^2}]$, is not
contained in the simulation domain, as seen from Figure \ref{fig:windprof}. Note that
the fast point is typically at very large distances in wind theory \citep{Bai_etal16}.
In practice, we have tested that containing fast magnetosonic point is not crucial to
wind kinematics\footnote{While working on the semi-analytical wind model of
\citet{Bai_etal16}, we have also solved time-dependent MHD wind equations along
prescribed poloidal field lines (unpublished). We find that the Alfv\'en point is quickly settled
even the flow is far from reaching the fast magnetosonic point, and the Alfv\'en radius
established early on is almost identical with the final steady-state solution.}

\begin{figure}
    \centering
    \includegraphics[width=90mm]{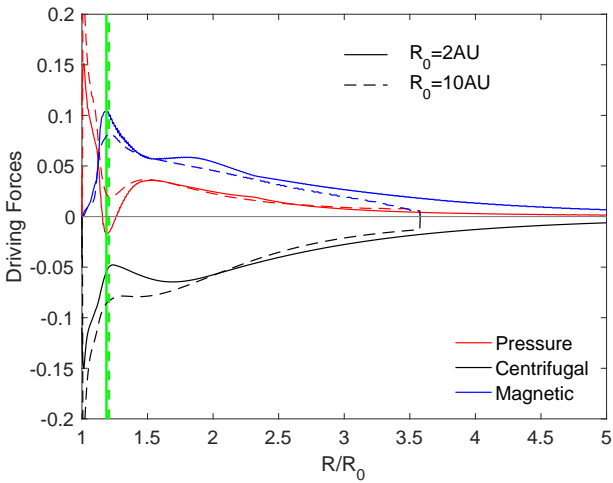}
  \caption{Decomposition of poloidal forces according to Equation (\ref{eq:decomp})
  along two representative field lines originating from $R_0=2$ AU (solid) and $10$
  AU (dashed), respectively, from the Hall-free simulation Fid0.
  Vertical dashed lines mark the location of the FUV front along
  the each field line.}\label{fig:force}
\end{figure}

By field line tracing at $R_0=2$ AU and $10$ AU, we find
$R_A/R_0\approx1.36-1.40$. According to Equation (\ref{eq:lever}), this implies that
$d\dot{M}_{\rm wind}/d\ln R\approx0.52-0.59$.
On the other hand, the actual mass loss rate appears to be a factor of $\sim3$ higher,
as seen from Figure \ref{fig:nh_acc}
This apparent discrepancy will be discussed in Section \ref{sssec:cmpsemi}.

The heavily loaded wind in our simulations is mainly driven by the toroidal magnetic
pressure gradient. To show this, we decompose the poloidal forces following Section
3.1.1 of \citet{Bai_etal16}
\begin{equation}\label{eq:decomp}
\frac{dv_p}{dt}=-\frac{1}{\rho}\frac{dp}{ds}
+\bigg(\frac{v_{\phi}^2}{R}\frac{dR}{ds}-\frac{d\Phi}{ds}\bigg)
-\frac{B_\phi}{4\pi\rho R}\frac{d(RB_\phi)}{ds}\ ,
\end{equation}
where the three terms correspond to thermal pressure gradient, the net centrifugal
force, and the Lorentz force from toroidal magnetic pressure gradient. Note that
we define the net centrifugal force as the excess of centrifugal force over gravitational
acceleration. We see in Figure \ref{fig:force} that this force is always negative, meaning
that corotation is far from being enforced to drive centrifugal fling, as in the conventional
\citet{BlandfordPayne82} picture.
Instead, acceleration is dominated by magnetic pressure gradient. This is because poloidal
fields in PPD winds are too weak to enforce corotation. They are thus wound up by
differential rotation, developing strong toroidal fields. We also see from
Figure \ref{fig:windprof} that $v_\phi$ falls off approximately as $R^{-1}$, and that
$|B_\phi|/B_p$ well exceeds $1$ in the wind zone. These results are
all consistent with the conclusions in \citet{Bai_etal16}. 

\begin{figure*}
    \centering
    \includegraphics[width=180mm]{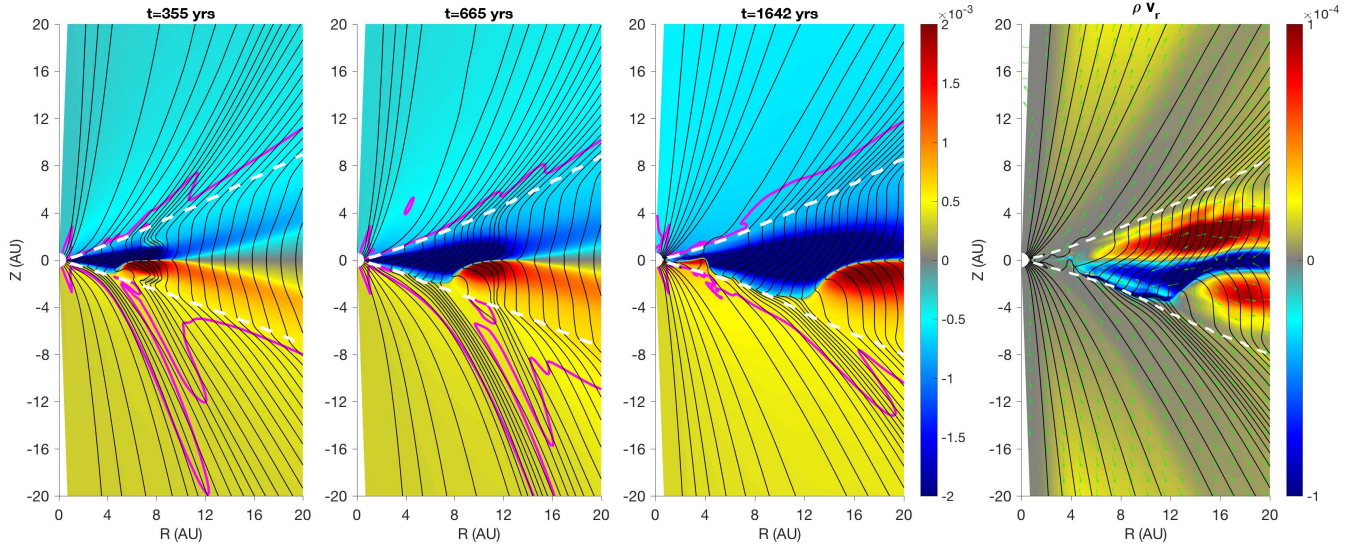}
  \caption{Same as Figure \ref{fig:nh_bevolve}, but for run Fid$+$ with all three non-ideal
  MHD terms included and vertical field aligned with disk rotation. Note that this run is
  restarted from the Hall-free run Fid0 at $t=3000\Omega_0^{-1}\sim222$ yr, after which
  the Hall term is turned on in an inside-out manner over a period of 5 local orbits. For
  reference, at $R=10$ and 20 AU, the Hall term is fully included after $380$ yrs and
  $669$ yrs. See Section \ref{sec:align} for more details.}\label{fig:hp_bevolve}
\end{figure*}

\subsubsection[]{Physics Behind Excessive Wind Mass Loss}\label{sssec:cmpsemi}

From Figure \ref{fig:nh_acc}, we estimate $d\dot{M}_{\rm wind}/\dot{M}_{\rm acc}\sim1.5$,
which translates to $R_A/R_0\sim1.15$ based on Equation (\ref{eq:lever}), indicating
extremely small lever arm. On the other hand, as mentioned earlier, when taking $R_0$
as the radius of the field origin at the midplane, we find $R_A/R_0\sim1.4$.
This apparent inconsistency is resolved by noting that when computing the lever arm,
$R_0$ should be defined as the radius of the wind base. More appropriately, it should be
taken to be the radius of the FUV front $R_{\rm FUV}$ (see Section \ref{ssec:nh_flow}).
In fact, from Figure \ref{fig:windprof}, we find exactly $R_A/R_{\rm FUV}\approx1.15$ at both
$R_0=2$ AU and $R_0=10$ AU.

Another apparent inconsistency arises when comparing the mass loss rate with semi-analytical
theory of \citet{Bai_etal16}. In our fiducial run, the poloidal Alfv\'en velocity and the sound speed
at the wind base are found to be around $0.1-0.15v_K(R_{\rm FUV})$, very close to the fiducial
parameter values adopted in \citet{Bai_etal16} (taken to be $0.1v_K$). However, although it was
pointed out there that PPD wind is heavily loaded, the predicted mass loss rate is about an order
of magnitude smaller than measured in our simulation.

Two factors contribute to the excessive mass loss rate in our simulation. First, poloidal field
strength $B_p$ appears to drop at a rate comparable to or faster than $R^{-2}$, as seen from
the bottom right panel of Figure \ref{fig:windprof}. In \citet{Bai_etal16}, it was found that the wind
lever arm is sensitive to how rapidly $B_p$
decreases with $R$. The fiducial model adopted there assumes $B_p\propto R^{-1}$ near the
disk, and transitions to $B_p\propto R^{-2}$ at larger distances, giving $R_A/R_0\sim2.3$. On
the other hand, assuming $B_p\propto R^{-2}$ dramatically reduces the lever arm with
$R_A/R_0\sim1.5$. Our simulation result suggests even faster decrease of $B_p$, which is
likely related to the fact that consecutive poloidal field lines are collimated at different levels,
where field lines originating from smaller radii are more collimated.

The second, and probably more important factor lies in the angular velocity of magnetic flux
surface $\omega$. We have shown that $\omega\approx\Omega_K(R_0)$. However, it is
more appropriate to normalize $\omega$ to $\Omega_K(R_{\rm FUV})$ (i.e., at the wind
base). For the two field lines shown in Figure \ref{fig:windprof}, we find
$\omega\approx1.25-1.3\Omega_K(R_{\rm FUV})$. This is substantially larger than the
range of values considered in \citet{Bai_etal16}, who considered the ratio in the range of
$0.95-1.05$. Despite the offset, a clear trend was identified that higher $\omega$ leads
to heavier mass loading (see their Figure 10).\footnote{While we can repeat the calculations
done in \citet{Bai_etal16} using higher $\omega$, we end up violating the assumptions made
there: the slow magnetosonic point is found to be well within the FUV front (where ideal MHD
no longer applies).}

We may further ask why $\omega$ deviates substantially from $\Omega_K$ at the wind base.
The reason is that the field is still largely anchored to the disk at radius $R_0$, thus rotating at
$\sim\Omega_K(R_0)$. In the disk upper layers (before reaching the FUV front) where X-rays
are the dominant ionization source, the coupling between gas and field is marginal, giving
$Am\sim1$. Such marginal coupling allows poloidal fields to bend, reaching the FUV front at a
larger radius $R_{\rm FUV}$. This makes the field lines rotate faster than the local Keplerian
speed, enhancing wind mass loading.

\section[]{Fiducial Simulation in the Aligned Case}\label{sec:align}

\begin{figure*}
    \centering
    \includegraphics[width=180mm]{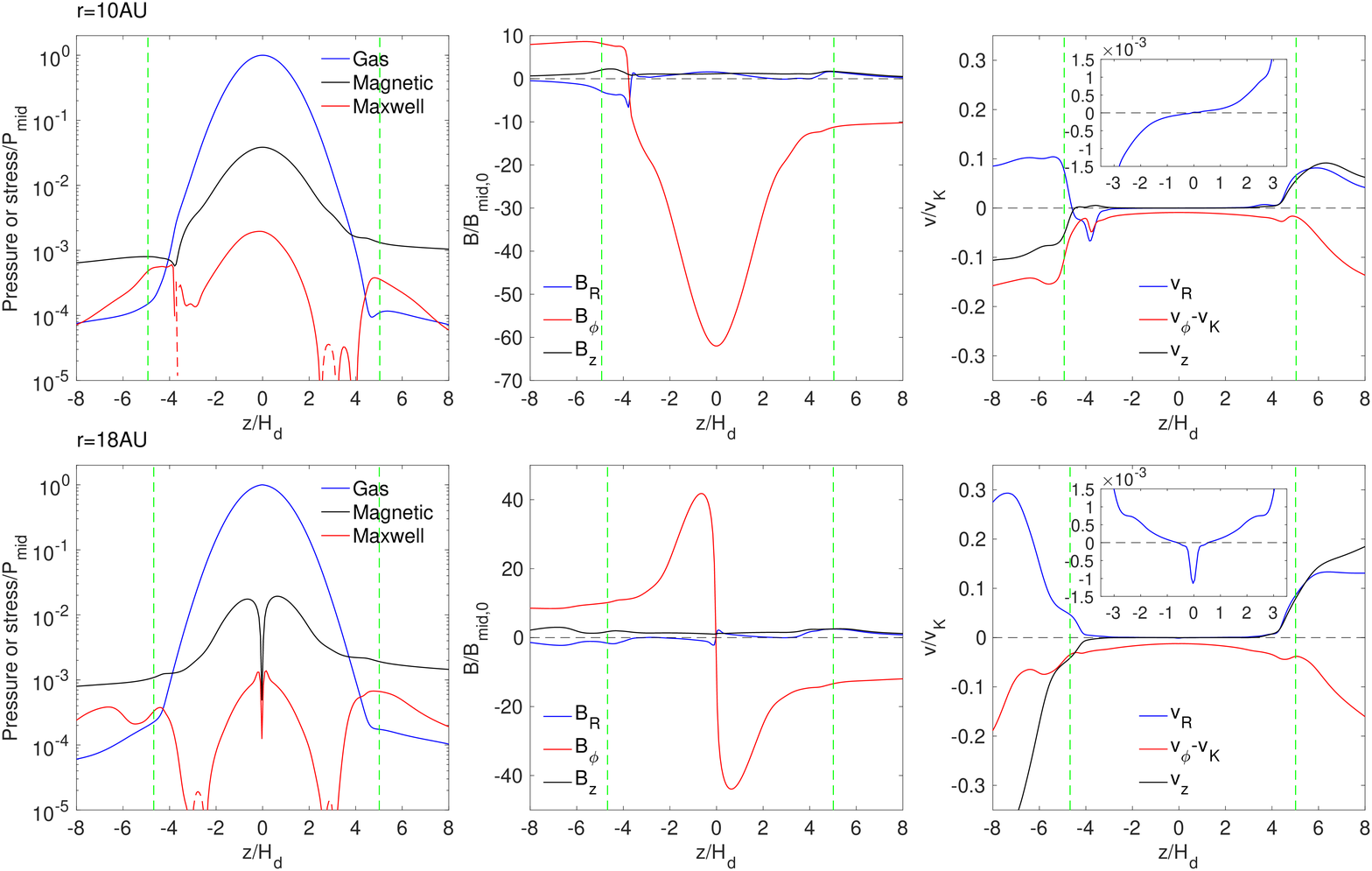}
  \caption{Same as Figure \ref{fig:nh_prof}, but for run Fid$+$, measured at around
  $t=21600\Omega_0^{-1}\sim1598$ yr, and the two chosen
  radii are $R=10$ AU and $R=18$ AU.}\label{fig:hp_prof}
\end{figure*}

Following the discussion in Appendix \ref{app:init}, for simulations including the Hall
effect with aligned field polarity, we restart from simulations from the Hall-free run
at a certain time $t_0$, and then turn on the Hall term in an inside-out manner.
This is motivated from evolutionary (i.e., disk formation) considerations detailed in
Appendix \ref{app:init}.
More specifically, we choose $t_0=3000\Omega_0^{-1}$ ($\sim222$ yr)\footnote{This
time is chosen so that it is sufficiently long for the inner region of the disk to be substantially
evolved, but still short enough so that asymmetric structures have not yet have time to
develop significantly according to Figure \ref{fig:nh_bevolve}.}, and turn on the Hall term
according to $[(t-t_0)/t_H]^4$ for $t<t_0+t_H$, where
$t_H$ is set to 5 local orbits. After $t=t_0+t_H$, the Hall term is completely included. The
simulation is run for sufficient amount time, measuring more than 10 local orbits at $R=20$ AU.

\subsection[]{Overall Evolution and Magnetic Field Configuration}\label{ssec:hp_bevolve}

In Figure \ref{fig:hp_bevolve}, we show the time evolution of magnetic field
configuration from run Fid$+$. Once the Hall term is turned on, horizontal
components of the magnetic field are quickly amplified due to the Hall-shear
instability (HSI, \citealp{Kunz08,Lesur_etal14,Bai14}). An MHD disk wind is
always launched in the presence of net poloidal field, maintaining a physical
geometry that poloidal field lines bend radially outward on both sides of the
disk. This physical field geometry requires toroidal field to change sign
across the disk \citep{BaiStone13b}. 

In the presence of HSI, we see that the disk can be clearly divided into two zones.
The inner zone shows prominent asymmetry, where the bulk disk is dominated by
toroidal field of a single sign (consistent with previous local simulations of
\citealp{Bai14,Lesur_etal14}), and toroidal field does not flip until reaching the disk
surface which roughly coincides with the FUV front. In the outer zone, symmetry
across the midplane is approximately preserved, where toroidal field flips (consistent
with some other local simulations of \citealp{Bai14,Bai15}, and recent global
simulations of \citealp{BaiStone17}). The two zones join
smoothly, and the location where toroidal field flips transitions from midplane
to surface over the range of a few AU.

The above phenomena are closely related to the development of the HSI. Globally,
the HSI is associated with radial transport of poloidal magnetic flux along the direction
of the Hall drift (or electron-ion drift in a dust-free gas), which stretches poloidal field
lines towards a radially-elongated configuration.
The radial field is then sheared by differential rotation in the disk to further amplify
the toroidal field \citep{BaiStone17}. The direction of the Hall-drift is along the electric
current, which is mainly due to the vertical gradient of toroidal field, whose growth
feeds back to the HSI. In the inner zone, we see that poloidal field lines
are highly inclined in the midplane region, which is the source of strong toroidal
field there. Similarly, poloidal field lines in the outer zone also have a significant
radial component, which strongly contrasts with the Hall-free run Fid0. The flip
of toroidal field is associated with a kink in poloidal field.

Generally speaking, the transition from the inner to the outer zone is related to the
transition from Hall-dominated to AD-dominated regime in the bulk disk.
This transition can already be traced from the Elsasser profiles shown in Figure
\ref{fig:elsprof} for run Fid0. For run Fid$+$ we find that around 12-20 AU, the Hall
term and AD have comparable strength in the midplane region, whereas the strength
of the Hall term drops rapidly towards the surface. In practice, there are more subtle
issues, which are
further discussed in Appendix \ref{app:symmetry} (e.g., comparison with the simulations
of \citealp{BaiStone17} which show fully symmetric solutions).

We also notice that in later stages of our simulation, there is a segregation of magnetic
flux in the inner few AU of the simulation box. This is partly related to the asymmetry in
the inner zone: poloidal field lines are highly inclined, and hence for the same
field line, it reaches the disk surface (where the wind is launched) at different radii in the
upper and lower sides of the disk. However, this asymmetry is broken by the presence of
the inner boundary. Once a field line is attached to the inner boundary, it loses disk
support and becomes isolated. In reality, however, it should be connected to some part
of the disk inside the inner boundary. In our simulations, we find that the inner boundary
gradually attracts magnetic flux from the lower side of the disk where the pinched
poloidal field line first reaches the inner boundary, and then the entire field line is accreted
to the inner boundary, building up magnetic flux there. While this flux segregation
phenomenon might be real to a certain extent,
it is clearly affected by the inner boundary, and hence is not very trustable.

In the rest of the discussion, we analyze the result at the end of our simulation, focusing
on regions characteristic of the asymmetric inner zone ($\sim8-12$ AU), and 
the more symmetric outer zone ($\sim16-20$ AU). We avoid regions affected by
the flux segregation phenomenon (within a few AU).

\begin{figure}
    \centering
    \includegraphics[width=90mm]{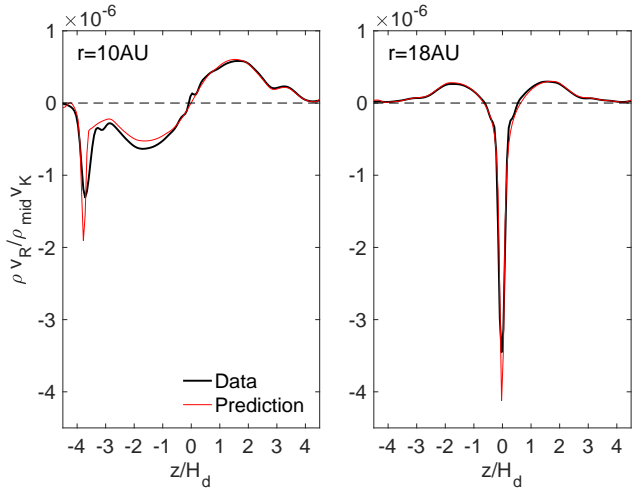}
  \caption{Vertical distribution of radial mass flux measured at two radii $R=10$
  AU (left) and $18$ AU (right) from run Fid$+$ (black) measured at around
  $t=21600\Omega_0^{-1}\sim1598$ yr, together with
  predicted mass flux based on Equation (\ref{eq:radflow}) in red.
  The horizontal dashed line marks zero
  mass flux to guide the eye. }\label{fig:hp_acc}
\end{figure}

\subsection[]{Angular Momentum Transport and Flow Structure}\label{ssec:hp_amt}

In this subsection, we discuss the mechanisms that drive disk angular momentum
transport in run Fid$+$, and the associated flow structure. 
In Figure \ref{fig:hp_prof}, we show the vertical profiles of major diagnostic
quantities at two representative radii $R=10$ AU and $18$ AU for the inner
and outer zones. We first discuss the results at these two locations separately
before analyzing the global diagnostics.

\subsubsection[]{Vertical Structure at the Inner Zone}

The inner zone is characterized by a strong toroidal magnetic field $B_\phi$ whose
strength peaks at the midplane.
Due to the HSI, we see from Figure \ref{fig:hp_prof}
that $B_\phi$ is amplified to $\sim60$ times the net vertical field. 
Note that even with such significant amplification, midplane magnetic pressure only
reaches a few percent of gas pressure (plasma $\beta\sim30$). This is consistent
with previous local simulations \citep{Bai14,Bai15}, and for given vertical field, the
amplification factor also depends on the assumed grain abundance (e.g., see
\citealp{Lesur_etal14,Simon_etal15b,XuBai16} where different grain abundances are adopted).

The vertical profile of $B_\phi$ is approximately symmetric about the
midplane before changing sign at one side of the disk surface. This configuration
largely determines the vertical profiles of the bulk flow velocity according to Equation
(\ref{eq:radflow}). In the left panel of Figure \ref{fig:hp_acc}, we find excellent agreement
between the radial mass flux measured in the simulation and expectation from Equation
(\ref{eq:radflow}). 

To elaborate, as the radial mass flux is largely determined by the vertical gradient of
$B_\phi$, it becomes clear that the bulk gas flows inward at one side of the disk, and
outward at the other side. Additional radial mass flux occurs at the position where
$B_\phi$ flips(at one side of the disk surface). The absolute mass fluxes in the
three regions are comparable (given the $B_\phi$ profile). On the other hand, because
the midplane region is much denser than the surface, radial flow velocity there is typically
small (as seen in the top right panel of Figure \ref{fig:hp_prof}), on the order of $1\%$ of
the sound speed. The flow velocity in the surface, on the other hand, is very significant,
and is on the order of the sound speed.	

We note that the net wind-driven accretion rate is determined by the difference in the wind
stress $-B_zB_\phi$ at the top and bottom disk surfaces. With $B_z$ approximately
constant within the disk, net accretion rate is largely set by $B_\phi$ at the surface.
While $|B_\phi|$ is a factor of several stronger in the midplane, it does not yield additional
mass flux, but leads to the radial inflow-outflow pattern whose mass fluxes largely cancel
each other. The net accretion flux mainly results from the surface layer where $B_\phi$
flips.

\begin{figure*}
    \centering
    \includegraphics[width=180mm]{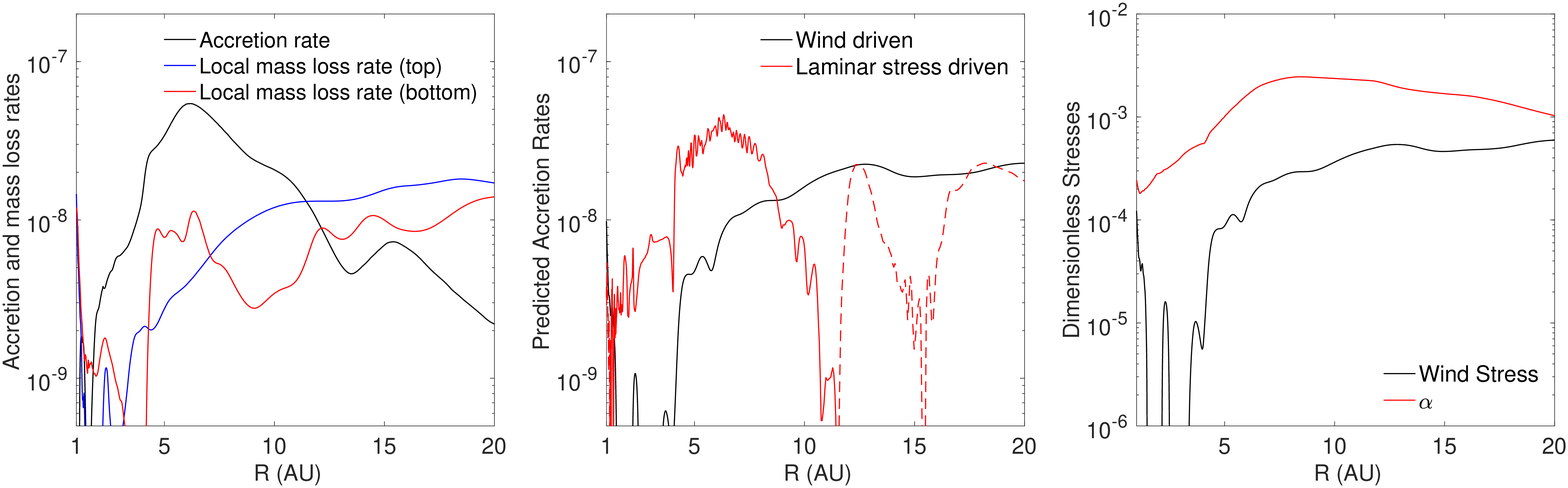}
  \caption{Left: radial profiles of accretion and outflow rates (per logarithmic radius)
  from run Fid$+$ around $t=21600\Omega_0^{-1}\sim1598$ yr.
  Note that mass outflow rates from the top and bottom sides of the
  disk are generally different. Middle: radial profiles of mass accretion rates derived from
  Equation (\ref{eq:accrate}), separating contributions from the wind (black) and the laminar
  Maxwell stress (viscously-driven, solid for accretion, dashed for ``decretion".).
  Right: radial profiles of normalized wind stress
  $|T_{z\phi}|/P_{\rm mid}$ and Shakura-Sunyaev $\alpha$ parameter [from Equation
  (\ref{eq:alpha})].}\label{fig:hp_radprof}
\end{figure*}

\begin{figure*}
    \centering
    \includegraphics[width=180mm]{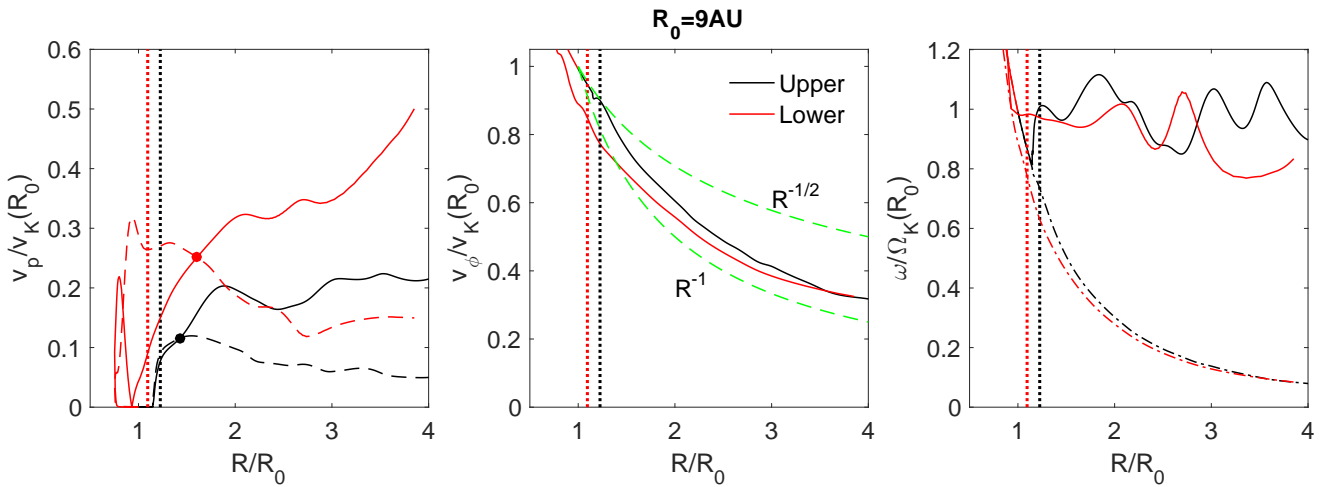}
  \caption{Wind properties in run Fid$+$, traced along a representative field line
  originating from $R_0=9$ AU at the midplane around
  $t=21600\Omega_0^{-1}\sim1598$ yr. The upper and lower sides of the
  line are marked by black and red, respectively.
  The three panels show poloidal velocity $v_p$, toroidal velocity $v_\phi$, and
  conserved quantity $\omega$ along the wind field lines.
  In the left panel, we further plot the poloidal Alfv\'en speed $v_{Ap}$ in dashed lines.
  Filled circles mark the locations of the Alfv\'en points. Vertical dashed lines mark
  the location of the FUV front.}\label{fig:hp_windprof}
\end{figure*}

\subsubsection[]{Vertical Structure at the Outer Zone}

The outer zone is characterized by an approximately symmetric field configuration
where $B_\phi$ flips at the midplane. Significant amplification of $B_\phi$
due to the HSI still occurs near the midplane, with amplification factor of up to $\sim40$
as seen in bottom middle panel of Figure \ref{fig:hp_prof} at the radius of 18 AU,
corresponding to a plasma $\beta$ of $\sim100$ at midplane.
The amplification factor will decrease towards larger radii as the Hall effect gets weaker.

The vertical profile of $B_\phi$ again leads to a very unusual radial flow structure,
mainly determined from Equation (\ref{eq:radflow}), as illustrated in the right panel of Figure
\ref{fig:hp_acc}. More specifically, the accretion flow is concentrated at the midplane
strong current layer where $B_\phi$ flips. On the other hand, because $|B_\phi|$ maximizes
right outside of the strong current layer, the drop in $|B_\phi|$ leads to radial outflows
in these regions, both above and below the midplane. Because these regions are not far
from the (dense) midplane, both outflow velocities, as well as the midplane accretion flow
velocity, are relatively small. At the radius of $R=18$ AU, they are on the order of $\sim1\%$
of the sound speed, as can be traced from the
bottom right panel of Figure \ref{fig:hp_prof}, where $c_s/v_K\approx0.09$.

The radial outflows above and below the midplane partially cancel the midplane
accretion mass flux. The net wind-driven accretion rate is again determined by the
difference in the wind stress $-B_zB_\phi$ at the top and bottom disk surfaces,
which amounts to roughly $20-30\%$ of the midplane accretion rate at $R=18$ AU.

\subsubsection[]{Radial Profiles of Accretion Rates}

To further analyze the mechanism of angular momentum transport, we
show in Figure \ref{fig:hp_radprof} the radial profiles of accretion and
outflow rates. We again treat the FUV front, marked as black dashed lines
in Figure \ref{fig:hp_bevolve}, as the wind base that separates the bulk
disk and the wind zone. We further show accretion rates derived from
Equation (\ref{eq:accrate}), separating contributions from the disk wind
and the laminar Maxwell stress.

We first analyze wind-driven accretion. As noted earlier, towards the
end of the simulation, there is a deficit of magnetic flux in the first few
AU of the domain, leading to very low wind-driven accretion rates. We
discard this region since it is likely related to the limitations of inner
boundary conditions. Wind-driven accretion rates reaches about
$2\times10^{-8}M_{\bigodot}$ yr$^{-1}$ beyond about 10 AU, with
normalized wind stress several times of $10^{-4}$. This is
comparable and slightly larger than the Hall-free case, which is also
consistent with local shearing-box simulation result \citep{Bai14,Bai15}.

The total accretion rate as seen in the left panel of Figure \ref{fig:hp_radprof},
however, differ significantly from wind-driven accretion rate. This is largely
owing to contributions from viscously-driven accretion from the laminar
Maxwell stress, as we discuss below.

From the rightmost panel of Figure \ref{fig:hp_radprof}, we see that the
$\alpha$ value peaks at around $7-10$ AU, which is right outside the
region deficit of magnetic flux. Towards larger radii, the $\alpha$ value
decreases, which is related to the fact that the Hall effect weakens. The
typical $\alpha$ value reaches a few times $10^{-3}$, which is consistent
with previous local simulations in this region \citep{Bai15,Simon_etal15b}.
Such $\alpha$ values (greater than $T_{z\phi}/P_{\rm mid}$ by a factor of
several) already suggests that viscously-driven accretion rate can be
significant compared with wind-driven accretion rate, as discussed in
previous local simulations \citep{Lesur_etal14,Bai14}.

\begin{figure*}
    \centering
    \includegraphics[width=180mm]{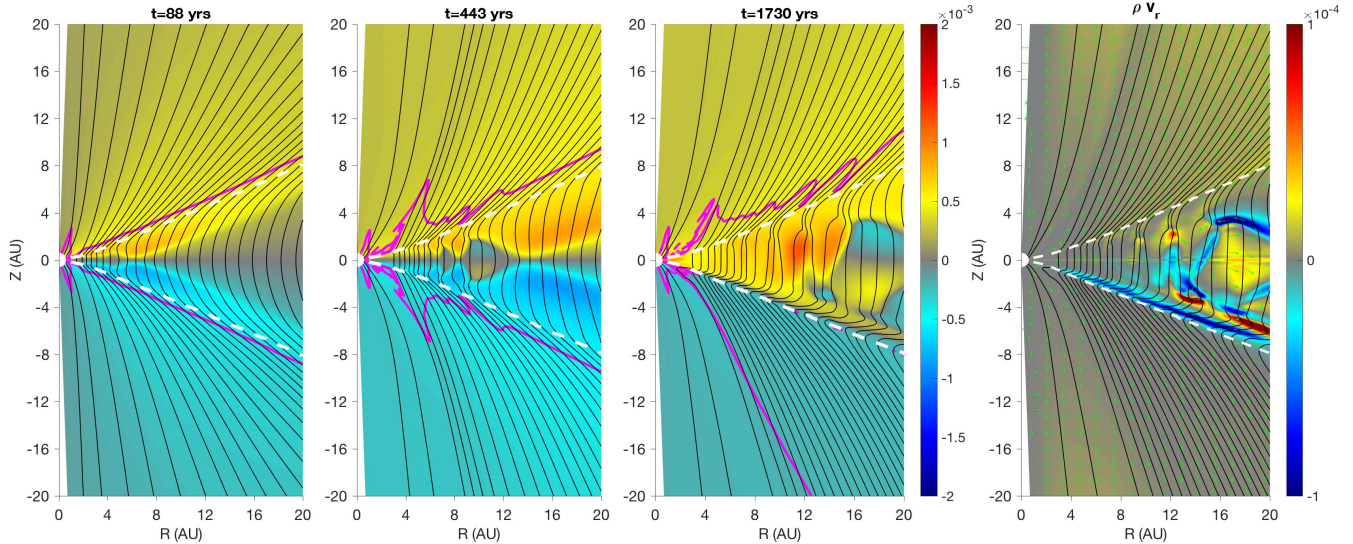}
  \caption{Same as Figure \ref{fig:nh_bevolve}, but for run Fid$-$ with all three
  non-ideal MHD terms included and vertical field anti-aligned with disk rotation.
  Note that the Hall effect is applied from the beginning.}\label{fig:hm_bevolve}
\end{figure*}

We note that viscously-driven accretion rate $\dot{M}_{{\rm acc},V}$ depends
on the radial gradient of $\alpha$ and other disk properties. For our adopted
radial surface density profile, we have
$\dot{M}_{{\rm acc},V}\propto d(\alpha\Sigma TR^2)/dR\propto d(\alpha R^{1/2})/dR$.
Therefore, if $\alpha$ decreases with radius more rapidly than $R^{-1/2}$,
the laminar stress would drive an ``decretion" flow instead of accretion.
This appears to be the case beyond $\sim10$ AU in our simulation, which
results in significant reduction of net accretion rate with increasing radius.
With this accretion rate profile, steady-state accretion is not possible,
although the duration of our simulations is too short to show significant
surface density evolution.

We emphasize that whether the laminar stress leads to accretion or decretion
is also affected by the overall density and temperature profiles, and our simulation
Fid$+$ can be considered as one realization at the given disk model and
magnetic flux distribution. While there can be many other possibilities,
one general trend likely holds. In regions where the Hall effect is important and
magnetic flux distribution is approximately uniform, $\alpha$ likely decreases with
radius, and hence reduces or even reverses viscously-driven accretion rate. Given
that the value of $\alpha$ is significant, the consequence of this effect on global disk
evolution can be very profound, which requires further investigations in the future.

\subsection[]{Wind Kinematics}

Overall, beyond a few AU (where magnetic flux has not evolved significantly),
the local mass outflow rate from our run Fid$+$ is comparable to that
from run Fid0, reaching $\sim10^{-8}M_{\bigodot}$ yr$^{-1}$, as seen from 
the left panel of Figure \ref{fig:hp_radprof}. For this particular run, mass outflow
rate there well exceeds accretion rate, due to the reduction of the latter from
viscously-driven decretion.

The asymmetry in the inner zone also leads to an asymmetry in the outflow rate:
the bottom side loses mass slower than the top side. In the outer zone near 20
AU where symmetry is retained, mass loss rate from the bottom side catches up
and approaches that from the top side. The wind kinematics in the outer zone
is similar to that discussed in the Hall-free case. Below we focus on the
wind kinematics of the asymmetric inner zone.

In Figure \ref{fig:hp_windprof}, we choose $R_0=9$ AU, and trace poloidal field
lines from the midplane both towards the upper and lower sides of the disk, and
measure various diagnostics along the field lines similar to those done in
Figure \ref{fig:windprof}. The asymmetry is already evident by looking at the
location of the Alfv\'en points, where the Alfv\'en radii are clearly larger at the
lower side of the disk. This can also be tracked directly from Figure
\ref{fig:hp_bevolve} over a broader range of radii. The smaller mass loss rate at
the bottom side of the disk is consistent with larger Alfv\'en radii, due to
Equation (\ref{eq:lever}). We can also see from the middle panel of Figure
\ref{fig:hp_windprof} that toroidal velocity in the wind drops faster in the upper
side than in the lower side, again consistent with the fact that wind from the
upper side is more heavily loaded, as discussed in \citet{Bai_etal16}.

\begin{figure*}
    \centering
    \includegraphics[width=180mm]{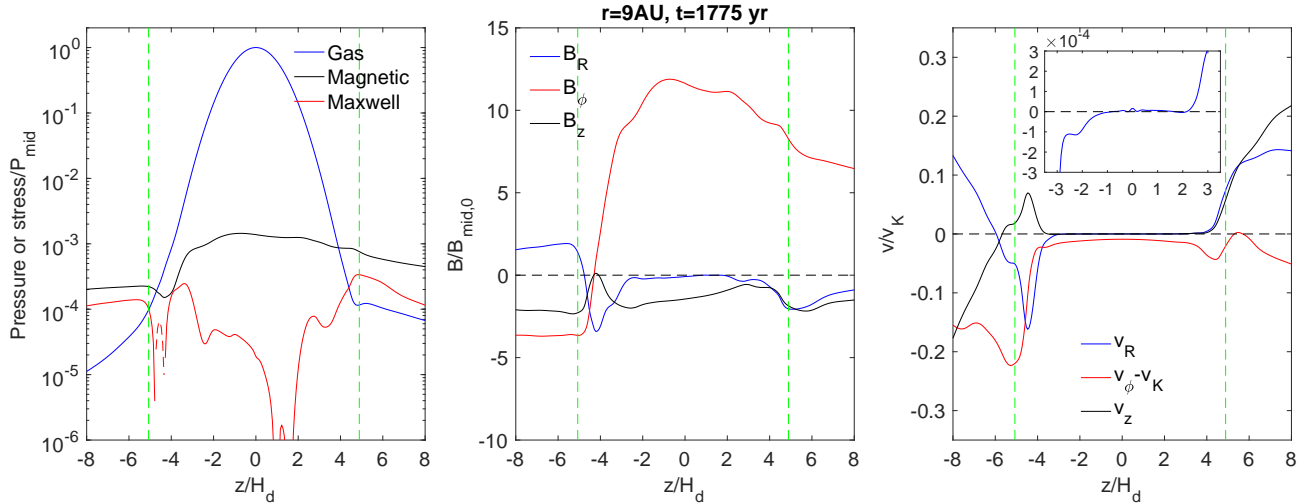}
  \caption{Same as Figure \ref{fig:nh_prof}, but for run Fid$-$, measured at the
  last simulation snapshot at $t=1775$ yrs. The analysis is done at cylindrical
  radius $R=9$ AU.}\label{fig:hm_prof}
\end{figure*}

To understand why wind in the lower side loses mass slower,
we note that because poloidal field lines at 9 AU are significantly inclined, the wind
bases at the top and bottom sides of the disk along this field line are located at
different cylindrical radii.  We see from the third panel that the angular velocity of the
field line is approximately the same at the upper and lower sides and is close to the
Keplerian frequency at radius $R_0$.\footnote{Towards the end of our simulation,
as regions in the first few AU down to the inner boundary is deficient of magnetic flux,
the system is less stable near the disk surface due to MRI-like activities (as the MRI
is a weak field instability), leading to perturbations to wind field lines at larger radii.
This is why $\omega$ in Figure \ref{fig:hp_windprof} in the wind zone is oscillating
around some constant value.} Because wind in the lower side of the disk is
launched from smaller radii, thus $\omega$ is smaller when normalized
to the local Keplerian frequency at the wind base. Similarly as discussed before,
this will lead to a larger Alfv\'en radius based on \citet{Bai_etal16}.

\section[]{Fiducial Simulation in the Anti-Aligned Case}\label{sec:antialign}

In the anti-aligned case, we find that the outcome of the simulation is
insensitive to initial conditions, and hence the simulations are performed
in the normal way as described in Section \ref{ssec:setup}. The simulation
is run for $24000\Omega_0^{-1}\approx1775$ yrs.

\subsection[]{Overall Evolution and Magnetic Field Configuration}

In Figure \ref{fig:hm_bevolve}, we show the time evolution of magnetic field
configuration from run Fid$-$. The initial stage of the evolution is very similar
to those found in \citet{BaiStone17}. Namely, instead of field amplification due to the
HSI, horizontal components of the field are reduced towards zero in the midplane
region. Overall, reflection symmetry across the midplane is preserved in this
initial stage, and magnetic flux is transported outward due to the Hall drift. The rate
of flux transport is rapid. By the time of $6000\Omega_0^{-1}\approx443$ yrs, the
inner $\sim3$ AU is largely depleted of magnetic flux (two other field lines in the
Figure are attached to the inner boundary), corresponding to a timescale
of $\sim100$ local orbits. This rate is comparable to (by order-of-magnitude) while a
factor of a few ($\sim3$) slower than the rate of flux transport reported in
\citet{BaiStone17} for the same given parameter ($\beta_0=10^{5}$).\footnote{As
discussed in \citet{BaiStone17}, the exact value of flux transport depends
on the diffusivity profile, and the profile adopted there was highly simplified.}

We also note that at $t=6000\Omega_0^{-1}\approx443$ yrs, in between
$R\sim5-12$ AU, the orientation of the poloidal field in the disk upper layers
points radially inward and outward in an oscillatory manner (with time and radius).
This leads to patches of toroidal fields of alternating polarities.
This is closely related to the phenomenon observed in earlier local simulations,
e.g., Figure 7 of \citet{Bai14} and Figure 9 of \citet{Bai15}, where more detailed
explanations were given. In brief, with anti-aligned vertical field, the disk becomes
more susceptible to the MRI in localized region in the disk where the Hall Elsasser
number is close to unity. In local simulations, such behavior may persist with time
(e.g., in the aforementioned figures), making it ambiguous to interpret its global
consequences. On the other hand, towards larger radii, toroidal field of one sign
eventually overwhelms, terminating the oscillatory behavior, as seen in Figure 11
of \citet{Bai15}. This is exactly what we observe in the global simulation Fid$-$,
illustrated in the third panel of Figure \ref{fig:hm_bevolve}. Once this sign of
$B_\phi$ is established, the inner region of the disk is quickly affected and
settle to the new asymmetric configuration. This pattern also propagates
outward, and by the end of our simulation, regions up to $\sim20$ AU are being
affected. Interestingly, under the new asymmetric field configuration, outward
transport of magnetic flux appears to be stalled, or at least significantly slowed
down.

Towards the end of the simulation, the overall field configuration is well established
within $R\sim12$ AU. While toroidal field is the dominant field component in the
bulk disk, its strength is only modest, in fact the mean field strength smaller than the
Hall-free case. Toroidal field then flips sharply at one (lower) side of the disk
slightly below the FUV front, creating a strong current layer. This is where most of the
accretion flow is concentrated (see the last panel), and the accretion flow is supersonic
(see next subsection).
In the upper side of the disk, we find that the poloidal field still show oscillatory behavior
(mainly beyond $\sim5$ AU) for same reason discussed earlier. This leads to some
slow motions in the midplane. Moreover, the deficit of magnetic flux within
$\sim3$ AU means that the supersonic surface accretion flow suddenly stops at $\sim3$
AU. In reality, it plunges into the inner regions, causes strong disturbances,
and also leads to rapid density variation in the outflows. This further affects the location of
FUV fronts at larger radii, leading to variabilities near the entire strong current layer. We
caution that because segregation of magnetic flux between the inner boundary and a few
AU is the main cause of this behavior, it may be subject to
the limitations in setting the inner boundary conditions discussed earlier.

\subsection[]{Angular Momentum Transport and Flow Structure}

\begin{figure*}
    \centering
    \includegraphics[width=180mm]{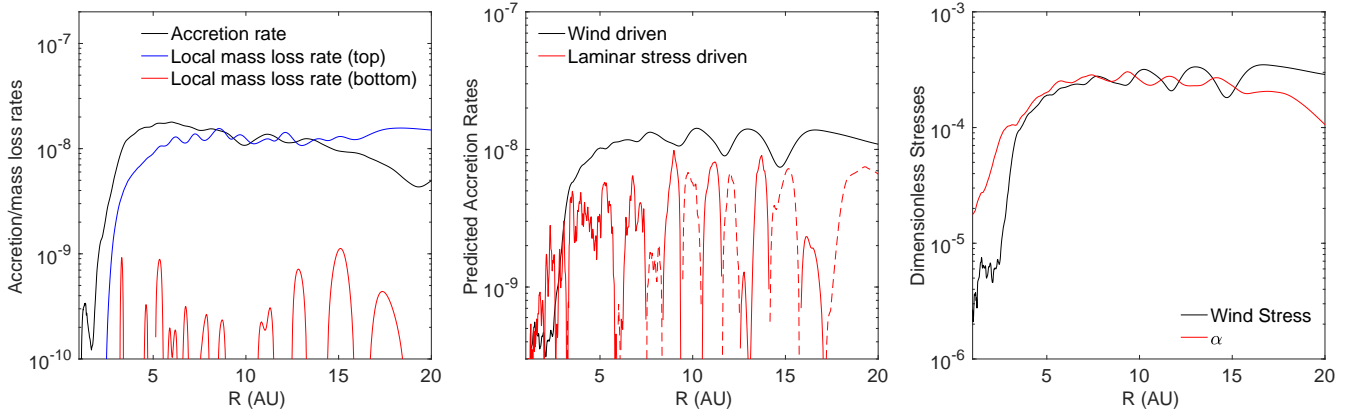}
  \caption{Same as Figure \ref{fig:hp_radprof}, but for run Fid$-$, measured around
  $t=1708$ yrs.}\label{fig:hm_radprof}
\end{figure*}

In this section, we first focus on a representative radius $R=9$ AU to analyze
disk vertical structure, and address the mechanism of angular momentum
transport in run Fid$-$.

\begin{figure*}
    \centering
    \includegraphics[width=180mm]{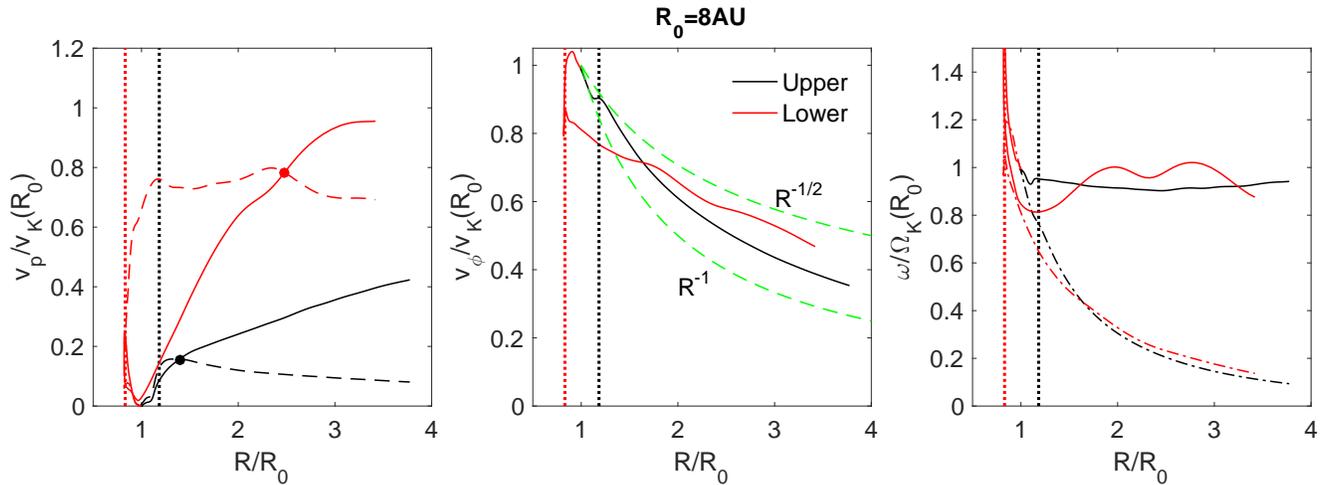}
  \caption{Same as Figure \ref{fig:hp_windprof}, but for run Fid$-$, with field
  lines traced from $R_0=8$ AU. Time averaging is performed between
  $18000-24000\Omega_0^{-1}\sim1331-1775$ yrs before tracing field lines.}\label{fig:hm_windprof}
\end{figure*}

\subsubsection[]{Vertical Structure}

In Figure \ref{fig:hm_prof}, we show the vertical profiles of major diagnostic
quantities at $R=9$ AU. Without amplifying the horizontal field by the HSI,
midplane field is much weaker than in run Fid$+$, with plasma $\beta\sim10^3$.
The disk maintains a small Maxwell stress $-B_RB_\phi$ in the bulk disk,
corresponding to $\alpha\sim10^{-4}$.

The flow structure is again mainly determined by the vertical gradient of
$B_\phi$ base on Equation (\ref{eq:radflow}). From the last panel of Figure
\ref{fig:hm_prof}, we see that radial velocity in the midplane region is largely
negligible. Essentially all the accretion flow is concentrated in the strong current
layer. We note that the location of the strong current layer is very close to the
FUV front where non-ideal MHD effects (dominated by AD) are greatly reduced,
making the layer thin (but is well resolved by more than $10$ cells).
The sharpness of the strong current layer leads to very rapid radial inflows,
where the velocity reaches $\sim15\%$ Keplerian speed. This is nearly twice
the sound speed in that region. This layer is likely unstable to the Kelvin-Helmholtz
instability, as found in surface current layers in some of \citet{Gressel_etal15}'s
Hall-free simulations. In our case, despite the flow being supersonic, it is not
straightforward to further discuss the stability of this layer due to the
disturbances in this layer discussed in the end of the previous subsection.

\subsubsection[]{Radial Profiles of Accretion Rates}

Similar as in the aligned case, we show in Figure \ref{fig:hm_radprof} the radial
profiles of accretion and outflow rates. Overall, the accretion rate is around
$10^{-8}M_{\bigodot}$ yr$^{-1}$. Accretion is primarily wind-driven, as can be
seen in the middle panel. Wiggles in the radial profiles of wind-driven accretion
rate/wind stress are related to the oscillatory behavior discussed earlier. Similarly,
the instantaneous $\alpha$ profile also exhibits wiggles. The corresponding
viscously-driven accretion rate shows accretion-decretion oscillations,
which time-averages to much smaller values than the absolute values
shown in the Figure. The typical $\alpha$ values reach $2-3\times10^{-4}$ over a
wide radial range, which is much smaller than in run Fid$+$, as expected.

\begin{figure*}
    \centering
    \includegraphics[width=150mm]{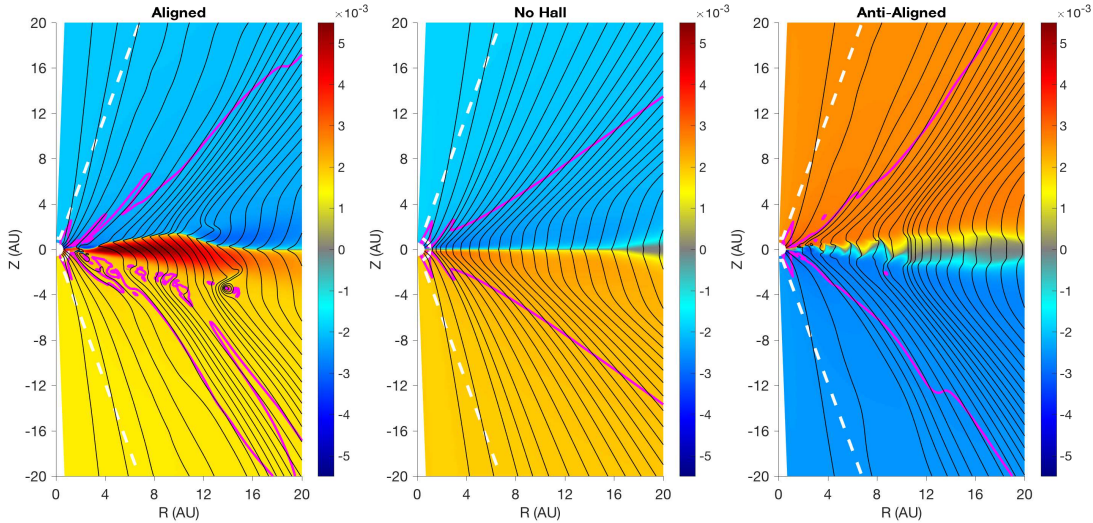}
  \caption{Magnetic field configuration at the last snapshot of our simulations with
  higher magnetization ($\beta_0=10^4$), run B4$+$ (left), B40 (middle) and
  B4$-$ (right).
  Black lines represent poloidal field lines represented as equally-spaced contours
  of constant magnetic flux, and color represents $RB_\phi$.
  Black dashed lines mark the FUV ionization front, red contours mark the location
  of Alfv\'en surface.}\label{fig:bevolve_b4}
\end{figure*}

\subsection[]{Wind Kinematics}

From Figure \ref{fig:hm_radprof}, we see there is a striking contrast between
mass outflow rates in the upper and lower sides of the disk. Mass loss
from the upper side is similar to those found in the grain-free case, whereas
from the lower side, the mass loss rate is about two orders of magnitude smaller!
This is related to the location of Alfv\'en surfaces, where we can identify from
the third panel of Figure \ref{fig:hm_bevolve} that at the lower side of the disk, it is
located at much larger distance than its counterpart at the upper side.

To understand this difference, we choose a characteristic radius $R_0=8$ AU,
and trace poloidal field lines from the midplane towards both the upper and lower
sides of the disk. In order to minimize the effect of disturbances to the field lines
in the bottom side of the disk, we have averaged the data between
$t=18000-24000\Omega_0^{-1}$ ($1331-1775$ yrs) before tracing the field lines.
In Figure \ref{fig:hm_windprof}, we show various diagnostics
along the field lines similarly as in Figure \ref{fig:hp_windprof}. Note that even
after time averaging, standard conservation laws are satisfied only approximately,
especially there are more deviations from the lower side of the wind (third panel).

We note that due to the large accretion velocity in the thin strong current layer,
field lines strongly pinch radially inward there. Because this layer lies right below
the FUV front, the wind base in the lower side of the disk has a radius that is
smaller than $R_0$. On the other hand, the wind base in the upper side
is located at a radius larger than $R_0$.  The relatively contrast between the two
wind base radii is large, amounting to a factor of $\sim1.4$ difference.
Note that the angular velocity of the field lines $\omega$ at the upper and lower
sides of the disk are approximately the same, and are comparable to Keplerian at
radius $R_0$ (third panel). This means that there is a large difference in $\omega$
when being normalized to the wind base radii at the upper and lower sides of the
field line. For similar reasons as discussed before, Alfv\'en radius must be larger at
the lower side of the disk based on \citet{Bai_etal16}, and hence much smaller
mass loss rate. 

\section[]{Parameter Study}\label{sec:param}

In this section, we briefly explore the role of net vertical field strength and the depth
of FUV penetration, only focusing on different features exhibited in these simulations.

\begin{figure*}
    \centering
    \includegraphics[width=180mm]{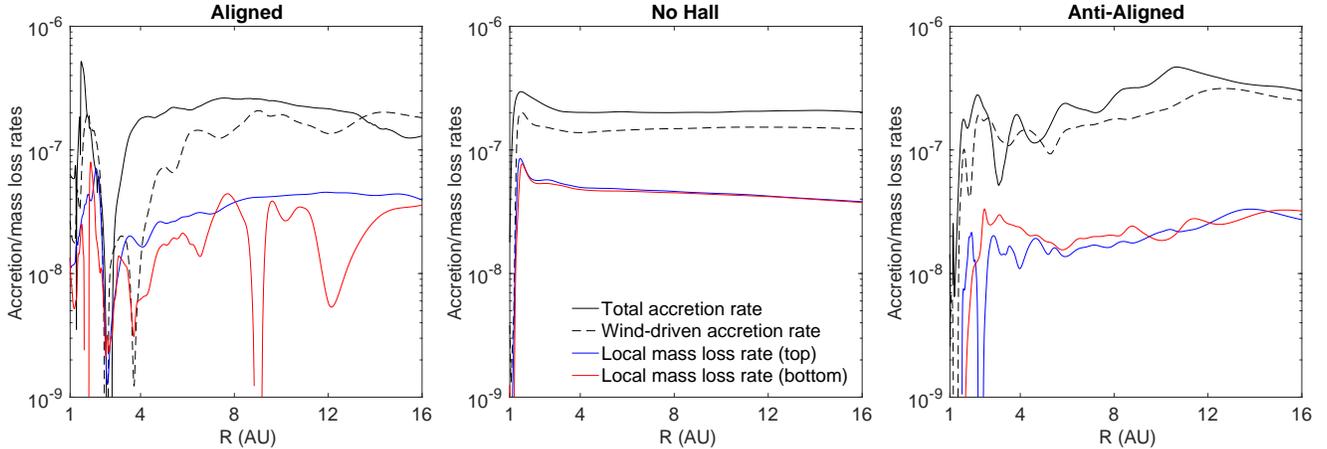}
  \caption{Radial profiles of accretion and outflow rates (per logarithmic radius) from
  our simulations with higher magnetization ($\beta_0=10^4$), run B4$+$ (left), B40
  (middle) and B4$-$ (right). The disk surface is fixed at $z=\pm5H_d$ in the
  calculation.}\label{fig:radprof_b4}
\end{figure*}

\subsection[]{Disk Magnetization}\label{ssec:beta}

With stronger magnetization ($\beta_0=10^4$), disk evolution proceeds much faster,
and hence it suffices for shorter run time. Same as before, runs B40 and B4$-$
are set up as described in Section \ref{ssec:setup}, while run B4$+$
is restarted from run B40 as described at the beginning of Section \ref{sec:align}.
In Figure \ref{fig:bevolve_b4}, we show the magnetic field configuration at the end of
each simulation. The radial profiles of accretion and mass loss rates from these runs
are shown in Figure \ref{fig:radprof_b4}.

\begin{figure}
    \centering
    \includegraphics[width=90mm]{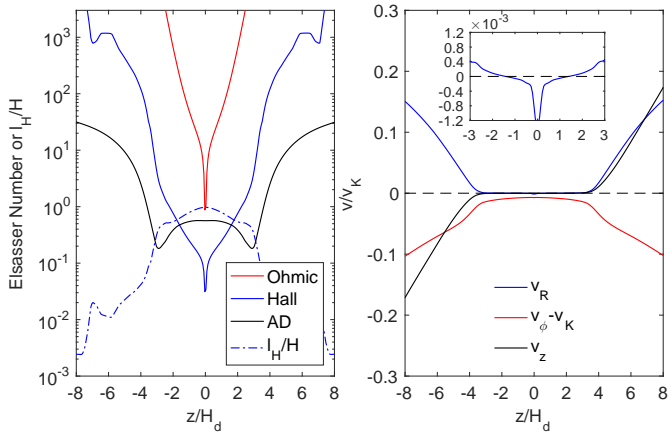}
  \caption{Vertical profiles of main diagnostic quantities from run B40
  at the last snapshot ($t=6000\Omega_0^{-1}\approx444$ yrs) at $R=6$ AU.
  The left panel shows the three non-ideal MHD Elsasser numbers
  (solid) and the normalized Hall length $l_H/H$ (dash-dotted).
  The right panel shows the three velocity components (Keplerian rotation
  subtracted), with the inset zooming in to highlight the accretion flow.}
  \label{fig:nohall_prof_b4}
\end{figure}

\subsubsection[]{The Hall-free Case}

We start by discussing the Hall-free run B40. Overall, we see from the middle panel
of Figure \ref{fig:bevolve_b4} that the field configuration is very
similar to run Fid0. With stronger net vertical field, the disk becomes more stable and
is fully symmetric about the midplane within the simulation domain. The toroidal
field flips at the midplane beyond $\sim4$ AU, where the accretion flow is concentrated.
Within that radius, toroidal field reduces to close to zero and the accretion flow splits
into two branches above and below the midplane.

With stronger magnetization, wind-driven accretion rate increases by about an order
of magnitude. Wind mass loss rate also increases, but by a smaller factor ($\sim3$).
Therefore, the wind becomes more lightly loaded, which is consistent
with the fact that Alfv\'en surface is located higher than that in run Fid0. This trend is
also consistent with expectations from semi-analytical theory \citep{Bai_etal16}.

The most interesting feature from this simulation is that because of higher wind mass
loss rate, the wind becomes denser and FUV radiation is almost completely shielded
from reaching the disk surface. This situation is distinct from earlier studies which attribute
efficient wind launching largely to FUV ionization (which brings the gas in the disk
surface layer to the ideal MHD regime) and associate the wind base with the FUV
front.

To further understand the nature of this wind, we show in Figure \ref{fig:nohall_prof_b4}
the vertical profiles of the Elsasser numbers and velocity components at a characteristic
radius of $R=6$ AU. In the wind zone, the dominant non-ideal MHD effect is AD, with its
Elsasser number $Am$ increases smoothly from $\sim1$ at about $z=\pm4H_d$ to
of order $\sim10$ or higher in the bulk wind zone. Without FUV, the wind column is
instead mainly ionized by the X-rays (the direct absorption component). The smooth
profile in $Am$ also leads to a smooth flow structure, where we see that the $v_\phi$
profile varies monotonically from midplane to surface, instead of having additional
peaks as in run Fid0 (see Figure \ref{fig:nh_prof}). 
We also find that in the wind zone, the conservation laws are not satisfied exactly, but
still approximately.

We caution that the simulations in this work are not designed to handle the situation
with FUV being shielded, and the physical condition in the wind zone can be far from
being realistic. In particular, without our prescribed heating beyond the FUV front, the
wind zone is even colder than the disk zone due to rapid expansion. This helps reduce
the recombination rate to achieve higher ionization level. In reality, heating
from X-rays (e.g., \citealp{Glassgold_etal04,Ercolano_etal09}) and ambipolar diffusion
\citep{Garcia_etal01} can be substantial. Moreover, equilibrium chemistry may no longer
apply, and even the prescription of X-ray ionization rate can become inaccurate in this
region. More careful calculations are needed to better understand the wind properties
in this regime \citep{Panoglou_etal12,WangGoodman17}. Overall, this simulation mainly
demonstrates that assisted by X-ray ionization, MHD disk winds can still be launched
when FUV radiation is shielded.

\begin{figure*}
    \centering
    \includegraphics[width=180mm]{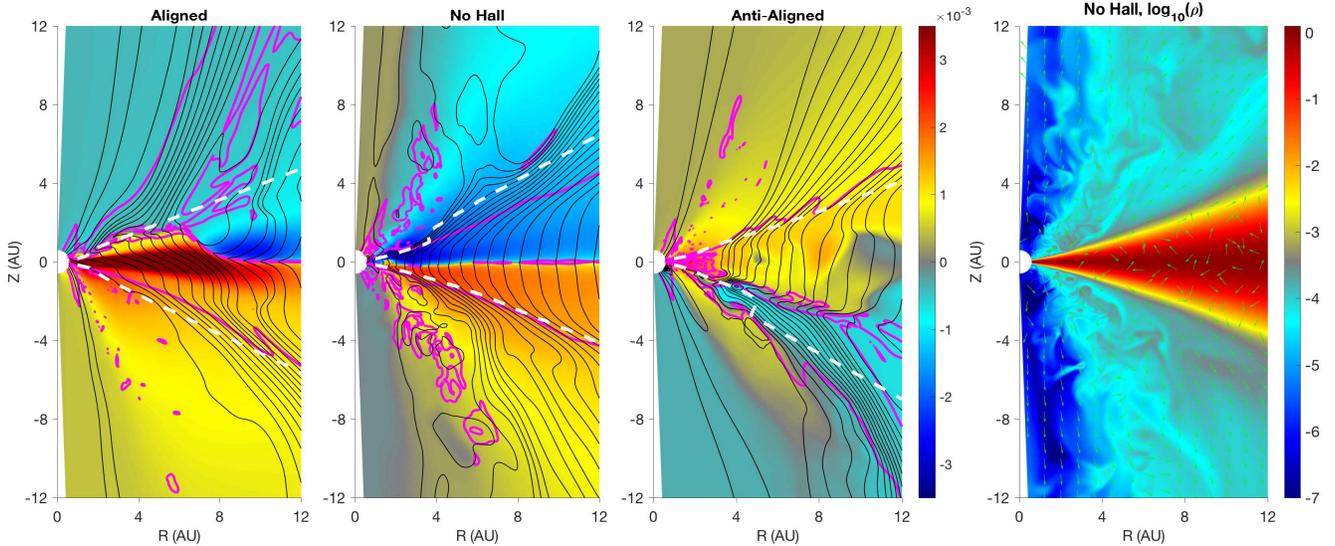}
  \caption{The first three panels are the same as Figure \ref{fig:bevolve_b4}, but for
  simulations with higher FUV penetration depth ($\Sigma_{\rm FUV}=0.3$g cm$^{-2}$),
  run FUV$+$ (left), FUV0 (middle) and FUV$-$ (right). Note that we have slightly
  time averaged the data spanning $60\Omega_0^{-1}$ in making these plots.
  The rightmost panel shows the density structure at the last snapshot (plotted as
  $\rho R^{q_D}$, no time average)
  of run FUV0. Green arrows indicate velocity vectors.}\label{fig:bevolve_FUV}
\end{figure*}

\subsubsection[]{The Aligned Case}

In the aligned case, we see from Figure \ref{fig:bevolve_b4} that again, the system
relaxes to a state with an asymmetric inner zone within $R\sim12$ AU, and a more
symmetric outer zone.\footnote{We note that despite using the more physically-motivated
initial conditions, we still find some localized regions possessing opposite toroidal
field from the HSI, forming some small localized poloidal field loops, though they
do not strongly affect the overall dynamics.}
The overall mass wind-driven accretion rate and mass loss rate
are comparable to the Hall-free run. With enhanced mass loss, FUV is again
shielded and wind launching occurs in the AD dominated surface layer. Partly because
of this, the vertical extent of the toroidal field patch (of single sign) in the asymmetric
inner zone is smaller (instead of extending to FUV front). Correspondingly, the wind
mass loss rates from the top and bottom sides of the disk are similar.

One general trend in the more strongly magnetized disks is that magnetic field
amplification factor through the HSI is smaller. At 10 AU, we recall that midplane
toroidal field in run Fid$+$ is amplified to $\sim60$ times the vertical field. In our
run B4$+$ and at the same radius, we measure that midplane toroidal field is only
amplified to $\sim25$ times the vertical field. Correspondingly, the laminar Maxwell stress
increases more slowly than the increase in the wind torque. As a result, wind-driven
accretion becomes more dominant, as can be seen in Figure \ref{fig:radprof_b4}.
Comparing with the fiducial run Fid$+$ (Figure \ref{fig:hp_radprof}), radial transport
of angular momentum clearly becomes less prominent in run B4$+$.

\subsubsection[]{The Anti-aligned Case}

With stronger net vertical field, the initial field evolution is similar to run Fid$-$,
with horizontal field reduced to close to zero at the midplane maintaining reflection
symmetry, and magnetic flux is rapidly transported outward. After about 1000 innermost
orbits ($\sim460$ years), the surface layer around a few AU becomes less stable,
leading to symmetry breaking with toroidal field of a single sign dominates, with
accretion proceeding at one side of the surface layer, again similar to run Fid$-$.
However, this state is very short lived. Due to significant mass loss, the FUV front is
again far from the disk surface, and wind launching proceeds in the AD-dominated
region. The system then settles to a state within $\sim10$ AU shown in the rightmost
panel of Figure \ref{fig:bevolve_b4}, where the strong current layer (and hence the
accretion flow) wiggles around the midplane region. With these wiggles, we find that
rapid loss of magnetic flux in this region is stalled. The region characterized by the
wiggled strong current layer slowly moves outward over time, and regions beyond
$\sim10$ AU are not yet affected by the end of the simulation (characterized by full
symmetry across the midplane with rapid outward flux transport).
While more detailed analysis is beyond the scope of this work, we note that in this
state, the overall disk dynamics is much more symmetric and mass loss rate from
the top and bottom sides of the disk are very similar.

\subsection[]{FUV Penetration Depth}\label{ssec:fuv}

\begin{figure*}
    \centering
    \includegraphics[width=180mm]{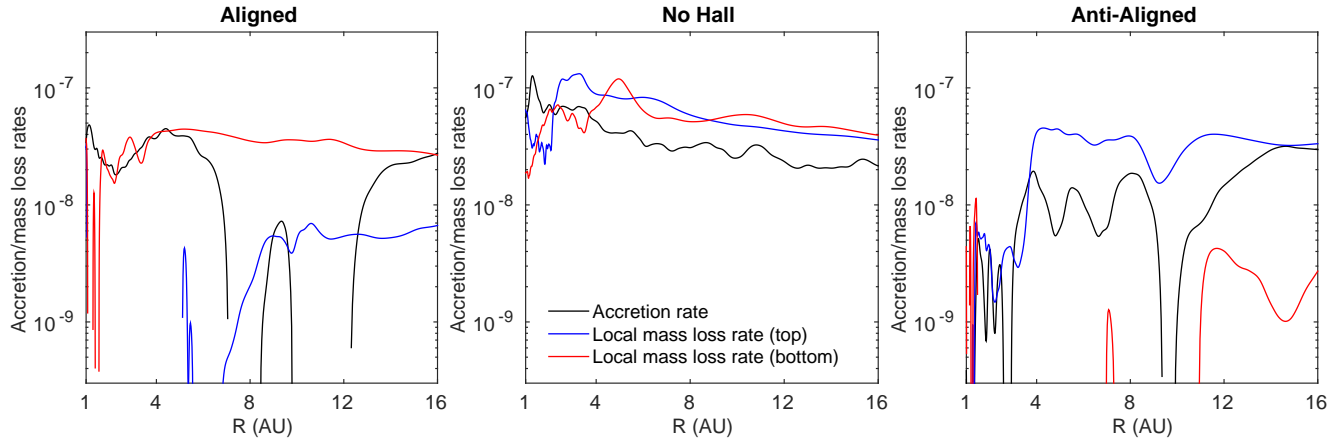}
  \caption{Same as Figure \ref{fig:radprof_b4}, but for simulations with
  higher FUV penetration depth ($\Sigma_{\rm FUV}=0.3$g cm$^{-2}$), run
  FUV$+$ (left), FUV0 (middle) and FUV$-$ (right).}\label{fig:radprof_FUV}
\end{figure*}

In Figure \ref{fig:bevolve_FUV}, we show the magnetic field configuration
at the end of each simulation with deeper FUV penetration
($\Sigma_{\rm FUV}=0.3$g cm$^{-2}$). The radial profiles of accretion and
mass loss rates from these runs are shown in Figure \ref{fig:radprof_FUV}.

Overall, the evolution of magnetic field configuration share many similarities
with the fiducial simulations. Without the Hall term, symmetry across the
midplane is roughly preserved, with toroidal field being the dominant field
component. With the Hall term in the aligned case, the HSI strongly amplifies
the horizontal field, leading to an asymmetric inner zone and a symmetric
outer zone with a smooth transition in between. In the anti-aligned case,
horizontal field is first reduced towards zero in the midplane, followed by
symmetry breaking. In the end, one sign of $B_\phi$ dominates the bulk disk,
with accretion flow concentrated at one side of the surface where this
$B_\phi$ flips.

With the Hall effect, the outflow from the top and bottom sides of the disk
show very significant asymmetry for both aligned and anti-aligned cases.
In particular, in the anti-aligned case, owing to the depletion of magnetic
flux between the inner boundary to about $R\sim4$ AU (which may be
unrealistic) and deep FUV penetration, the bottom side of the disk surface
where $B_\phi$ flips shows complex evolution (likely due to the MRI), and
starts to tangle the poloidal magnetic flux by the end of our simulation (and
the measured ``outflow" rate becomes negative in some regions).
Further investigation of this issue is desirable but beyond the scope of this
work. The discussion of outflows in the next subsection, on the other hand,
remains applicable to the top side of the disk for run FUV$-$.

\subsubsection[]{Turbulent Outflow and Shielding}

One important influence of larger $\Sigma_{\rm FUV}$ is that it makes the
outflow turbulent and drives much more significant mass loss. Three main features
are worth discussing.

First of all, we find that in all three runs, the disk surface layer becomes unstable
within radius $r\sim2$ AU, and launches episodic outflows. This is because for
given vertical field strength, the well ionized disk surface layer near the FUV front
becomes less magnetized, making it more susceptible to the MRI. As an example,
the rightmost panel of Figure \ref{fig:bevolve_FUV} shows the density and velocity
structure from the last snapshot of the FUV0 run, which are clearly indicative of
vigorous and turbulent outflow activities originating from the inner regions.

Second, beyond $\sim3$ AU, we find that the disk is relatively stable. This is mainly
because that the unsteady outflow launched from smaller radii is so dense that it
substantially shields the FUV radiation. In fact, the FUV front in run FUV0 at the
distance of a few AU is located higher than that in run Fid0, despite that the former
has much larger $\Sigma_{\rm FUV}$! As a result, the disk surface layer is stable
against the MRI.

Third, the bulk wind-driven accretion rate and wind mass loss rate in all three runs
all exceed those in the fiducial case. Moreover, the ratio of mass loss to accretion
rate is higher. This result first appears natural, and agrees with semi-analytical
studies as the natural outcome of deeper FUV penetration \citep{Bai_etal16}.
This argument likely applies only in the innermost $\sim2$ AU (although the wind is
episodic in this region),
but does not apply to the outer region because the FUV front is located even higher
than in the fiducial runs. The main reason for the enhanced accretion and mass
loss rates is that the ram pressure from the heavily loaded episodic inner wind
pushes the poloidal field lines and bends them further. This is similar to reducing
the $\theta$ angle in the \citet{Bai_etal16} wind model. Although this parameter
was only very briefly explored there, the trend is that more inclined field tends to
develop stronger toroidal field, leading to both enhanced accretion and outflow
rates, with the latter being more pronounced. This is consistent with what we
observe in Figures \ref{fig:bevolve_FUV} and \ref{fig:radprof_FUV}.

Moreover, we notice that in many cases, the Alfv\'en radius is located within
the FUV front. Namely, the development of the wind proceeds in the presence
of strong non-ideal MHD effect. We have already discussed this phenomenon
in the previous subsection. It again adds more complications to the dynamics
of the system. On the other hand, we have also seen that semi-analytical theory
is still useful that helps interpret the basic trend.

Finally, we caution that because part of the inner disk is MRI unstable, our 2D
simulations are unable to fully characterize the flow properties, and hence
the properties of the episodic winds launched from these regions. Full 3D
investigations are necessary to resolve the gas dynamics more self-consistently.

\begin{figure*}
    \centering
    \includegraphics[width=180mm]{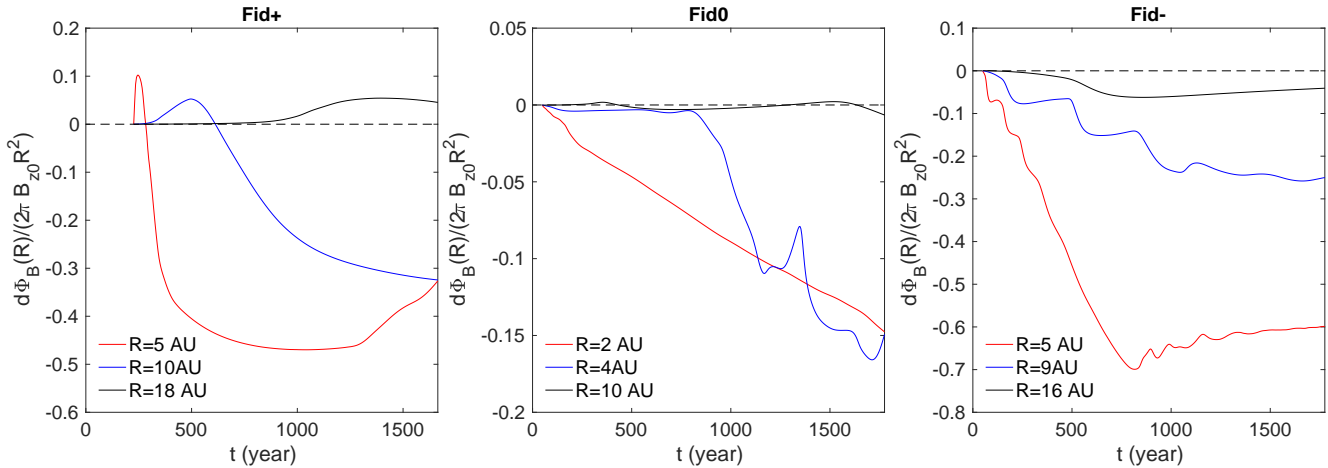}
  \caption{The amount of magnetic flux that goes through the midplane
  region of the disk at certain radii $R$ (marked in legends in each panel) as a function
  of time for the three fiducial simulations shown in the three panels. The magnetic flux
  is normalized $2\pi B_{z0}R^2$ where $B_{z0}$ is the initial vertical field.}\label{fig:flux_trans}
\end{figure*}

\section[]{Discussion}\label{sec:discussion}

\subsection[]{Magnetic Flux Evolution}

Given the very pronounced differences in simulations with different initial
poloidal field strengths, it is clear that the overall disk evolution is largely
controlled by the strength and radial distribution of poloidal magnetic flux.
Following from Section \ref{ssec:issues}, we now discuss
the global evolution of magnetic flux based on our fiducial simulations.
For each run, we compute the magnetic flux function $\Phi_{B,{\rm mid}}$
from Equation (\ref{eq:PhiB}), and follow its evolution at three representative
radii (different set of radii for different runs, since the dynamics and field
configuration are different in each case).
In Figure \ref{fig:flux_trans}, we show the time evolution of
$d\Phi_B(t)\equiv\Phi_{B,{\rm mid}}(r,t)-\Phi_{B,{\rm mid}}(r,t_0)$,
the amount of magnetic flux that has been transported through the chosen
locations since time $t_0$ (time when magnetic flux is introduced, or when the
Hall term is turned on in the case of run Fid$+$). The values are normalized by
$2\pi B_{z0}R^2$, where $B_{z0}$ is the initial vertical field strength, so that
one can easily estimate the timescale of flux transport.

In the Hall-free run Fid0, we
see that at $R\sim2$ AU, magnetic flux is consistently transported.
From the figure, we estimate the timescale of flux transport to be
$\sim10^4$ yr, translating to a speed of
$v_B\sim3\times10^{-5}v_K$. This is more than an order of magnitude slower
than the rate found in \citet{BaiStone17}. As speculated there, the rate of flux
transport is sensitive to the vertical diffusivity profile. On the other hand, the
timescale of $10^4$ year is still too short compared with disk lifetime.
At $R=4$ AU, flux is also transported outwards despite some irregularities,
which is related to the secular evolution of toroidal field patches in Figure
\ref{fig:nh_bevolve} that precludes accurate measurements of flux transport rate.
At $R=10$ AU, on the other hand, we do not find obvious signs of outward flux
transport (despite small variations) within $\sim1500$ yrs, amounting to
$\sim50$ local orbits of evolution.
If the same rate of flux transport measured at $\sim2$ AU applies here, we
would expect $d\Phi_B/(2\pi B_{z0}R^2)\sim0.01$ by $\sim1200$ yrs, yet
this is not achieved. Therefore, the rate of flux transport at larger radii is even
slower.

In the aligned case, run Fid$+$, we see that at all three radii, magnetic flux follow
a pattern of being transported inward first ($d\Phi_B>0$), followed by outward
transport. This is consistent with the findings in \citet{BaiStone17}, as a result
of the HSI, followed by outward diffusion (at $R=18$ AU, outward diffusion just
starts to develop by the end of the simulation). There is some anomalous trend
at $R=5$ AU towards later stages, where we have discussed in Section
\ref{ssec:hp_bevolve} that might be an artifact due to inner boundary conditions.
None of the three representative radii have achieved a quasi-steady state in
magnetic flux evolution, which again makes it difficult to assess the overall rate
of flux transport.

In the anti-aligned case, run Fid$-$, we see that at the beginning, magnetic flux
is systematically transported outward at all radii, at a rate that is several times
faster than the Hall-free case. This is overall consistent with the findings in
\citet{BaiStone17}. However, after the symmetry across the midplane is broken,
a sudden change in the rate of transport is induced. At $R=5$ AU, we find that
flux is even transported slowly inward, whereas in between at $R=9$ AU, flux
evolution comes to a stall. We note that in this field configuration, the vertical
profile of toroidal field is mostly flat, and hence substantially reduces the radial
Hall-drift (proportional to $\pa B_\phi/\pa z$). This is likely the main reason for
the reduction and even reversal in flux transport. More detailed analysis about
the direction and steady-state rate of flux transport is beyond the scope of this
work, but is worth pursuing in the near future.

Overall, we have seen that while the initial stages of magnetic flux evolution in
our simulations are similar to those found in \citet{BaiStone17}, more complex
behaviors are found as the system develops more complex field configurations
and flow structures, resulting from more realistic prescriptions of disk microphysics.

\subsection[]{Comparison with Other Works}\label{ssec:compare}

As this work was in preparation, \citet{Bethune_etal17} (hereafter BLF17) conducted
similar types of simulations of PPDs, and reported a variety of phenomena related
to disk angular momentum transport, wind launching, etc.
In methodology, there are two major differences between our simulations and theirs.
First, the simulation domain in BLF17 extends to about $\pm60^\circ$ above/below the
midplane, truncating at least part of the wind launched from the disk, as well as some of
the magnetic flux in the simulation box. Our simulations do not suffer from this limitation.
Second, our simulation domain covers a factor of 100 in radius as opposed to 10,
which allows us to comfortably follow the launching and propagation of disk winds.
There are several other differences at implementation level. To list a few,
our simulations use more realistic tabulated magnetic diffusivities based on
a complex chemical reaction network containing dust grains, as opposed to analytical
diffusivity prescriptions mimicking the grain-free case in BLF17. 
We consider a flaring disk geometry, and use ray-tracing to estimate the radial and
vertical disk column densities, whereas BLF17 considered flat disks (constant $H_d/R$),
and did not account for the radial column. The treatment of the transition from the disk zone
to the atmosphere is very different. The transition in BLF17 is located at prescribed and
constant latitudes, and is generally closer to midplane (by more than one scale height) than
ours, leading to more unstable surface layer and more significant mass loss. Moreover,
most of the BLF17 simulations are significantly more strongly magnetized, leading to
accretion rates that are at least an order of magnitude higher than our fiducial runs.

Compared with a large variety of behaviors found in BLF17, we find a much more unified
set of behaviors that are unique to the aligned and anti-aligned field polarities. In particular,
BLF17 found ``non-accreting" cases where $B_\phi$ vanishes in the disk corona region,
with meridional circulations but no wind-driven accretion. We do not observe this behavior: wind
is always found leading to net accretion, on top of which there are meridional flows in the aligned
cases as explained in Section \ref{ssec:hp_amt}. We speculate that the non-accreting solutions
in BLF17 may be related to their over-constraining domain size. With much larger
dynamical range and longer simulations, we have also identified and clarified regimes
where the disk/wind structure become symmetric or asymmetric with respect to the midplane.
Moreover, the diversity of behaviors in BLF17 is also likely related to the sensitivity to initial
conditions in the aligned case, as we discuss in Appendix \ref{app:init}. By mimicking conditions
of disk formation, we have (at least partially) avoided this problem.

BLF17 also considered ``cold" wind and ``warm" wind, where the wind region is heated
to very different temperatures. Most of the behaviors are consistent with the semi-analytical
magneto-thermal wind framework of \citet{Bai_etal16}. In particular, we note that keeping all
other parameters fixed, while the ``warm" wind simulations lead to much stronger acceleration
due to thermal driving, the wind mass loading (or mass loss rate) remains similar to the
``cold" wind. In other words, wind mass loss rate is largely determined by the conditions at the
wind launching region, but not subsequent thermal accelerations. BLF17 described the wind
from several of their warm simulations with different properties as ``magneto-thermal". In
Appendix \ref{app:interpwind}, we aim to systematically clarify the nomenclature on PPD
winds and the corresponding phenomenology.

Finally, we comment that we do not observe the magnetic flux concentrations and zonal flows
in our 2D simulations. The phenomenon identified and explained in BLF17 is found in highly
strongly magnetized (and hence accretion rate is several orders of magnitude higher) disks
with net vertical flux $\beta_0=10^2$, a regime not explored in our simulations. We also
comment that the mechanism is also likely related to the sharpening of flux concentration
observed in \citet{Bai15}.

\subsection[]{Implications for Planet Formation and Disk Evolution}

The most important implication of this work on planet formation is from the
complex flow structures.
In particular, in simulations with aligned field polarity,
the presence of both accretion and decretion flows at different heights in the bulk
disk is completely unexpected based on conventional models (especially viscous
evolution models) of accretion disks. It poses very interesting questions on how
it would affect the transport of solids, and subsequent stages of planet formation.
The systematic outward motion in a substantial fraction of the bulk disk may be an
important source for large-scale mixing. Evidence for such radial mixing
in the solar system has been mounting, especially based on the findings of
crystalline silicates in comets  (e.g., \citealp{Bockelee_etal02} and references therein),
and more directly from samples collected from comet 81P/Wild 2
\citep{Brownlee_etal06,Nakamura_etal08}. Similar evidence has been found in
nearby PPDs \citep{vanBoekel_etal04,Watson_etal09}. Viscous diffusion with
large-scale radial flows have been commonly involved to explain such large-scale
mixing (e.g., \citealp{KellerGail04,Ciesla09,HughesArmitage10}), as well as
variations in various isotopic ratios in the solar system, such as the D/H ratio in
water \citep{JacquetRobert13,Yang_etal13,Albertsson_etal14}.
Our simulation results offer a first-principle demonstration of the large-scale
flow structure that differ substantially from the conventional picture.
With typical radial flow velocity of the order $\sim1\%$ sound speed, and given
that the sub-Keplerian velocity in the inner regions of PPDs is around $\sim5\%$
of sound speed, this means that decretion flow in the bulk disk can overcome radial
drift for particle Stokes numbers ${\rm St}\lesssim0.1$. This flow thus has the potential
to transport mm-sized particles to $\sim30-40$ AU scale (if our results can be
generalized to outer radii) in a standard MMSN disk. More detailed calculations are
necessary to further demonstrate its feasibility.

Global evolution of PPDs is determined by the transport of angular momentum, both
radially and vertically, as well as mass loss.  Our simulations with aligned field polarity
also show dramatic radial variations in accretion rate due to significant contribution
from the laminar Maxwell stress as a result of the HSI (whose rate and even
direction depends on the radial gradient of the stress). This fact, together with
significant mass loss rate, implies that global disk evolution is highly complex, and it is
unclear whether a steady state can ever be achieved. Uncertainties in magnetic flux
transport and evolution discussed earlier add further complications.

The discussions above mainly focused on the aligned case. PPD gas dynamics in the
anti-aligned case shows completely different behaviors. It is thus very likely that
planet formation takes very different pathways in these two cases, though the details
remain to be filled in upon better understandings of long-term disk evolution (but see
a toy model by \citealp{Simon16}).

Finally, we speculate that the solar nebular was once threaded by poloidal field with
aligned polarity, for two reasons. First, large-scale outward radial flows are only
present in simulations in the aligned case. While outward transport is also possible
in the anti-aligned case through diffusion, the weak level of turbulence is unlikely
to lead to efficient large-scale mixing. Second, recent paleomagnetic measurement
of the Semarkona meteorite revealed a strong magnetic field of $\sim0.5$ Gauss
\citep{Fu_etal14},\footnote{This value should correspond to nebular field strength,
unless the Semarkona chondrules are formed by nebular shocks. However, very
recent modeling work shows that even in such shocks, the level of
ionization in the shock downstream is not high enough to compress magnetic field
along with gas (Mai, Desch \& Boley, in preparation), thus the measured paleointensity
directly records nebular field strength.}
presumably corresponding to the asteroid belt region in the midplane. We note that
that in the aligned case, radial transport of angular momentum is comparable to wind
contributions. For typical accretion rate of $10^{-8}M_{\bigodot}$ yr$^{-1}$, $\sim0.5$G
is exactly the expected field strength at $\sim2-3$ AU scale
from radial transport of angular momentum (see \citealp{Wardle07,BaiGoodman09}).
Comparing Figures \ref{fig:hp_bevolve} and \ref{fig:hm_bevolve}, we see that the
midplane field in the anti-aligned case is typically a factor of $4-5$ times lower,
inconsistent with the paleomagnetic measurement.

\subsection[]{Connection to Disk Observables}

One important prediction from this work is that disk winds from the inner
region of PPDs are likely asymmetric between the two sides. While wind
signatures seem to be ubiquitous among T Tauri disks
\citep{Hartigan_etal95,Natta_etal14,Simon_etal16} based on optical-infrared
forbidden line (blue-shifted) observations, these observations typically see winds 
only from one side, and the observations themselves already bare large uncertainties in
constraining wind kinematics. Jets from young-stellar-objects
often show asymmetric signatures between the jet and counter-jet (e.g.,
\citealp{Hirth_etal94,Woitas_etal02,HartiganHillenbrand09,Liu_etal12}), although it
is less clear whether such asymmetry extends to the lower-velocity wind components.
Recently, ALMA has revealed large-scale molecular outflows in several sources
(e.g., \citealp{Klaassen_etal13,Salyk_etal14,Bjerkeli_etal16}), which all show complex
spatial and velocity structures, some of which are only one-sided. Environmental effects
(e.g., envelope, foreground, tidal interaction with binary) may be important contributing
factors, but it is unclear whether some are caused by intrinsic asymmetry during the
wind launching processes. Very encouragingly, \citet{Klaassen_etal16} derived the
kinematics of disk winds from the HL Tau disk (despite the systematics), and found
dramatically differences (by $\sim$an order of magnitude) between the redshifted and
blueshifted sides.

The complex flow structures in the bulk disk found in our aligned simulations reach
systematic radial velocities of up to a few percent of the sound speed. We note that
careful modeling of ALMA data has already enabled level of turbulence to be constrained
at a comparable precision (Flaherty et al. submitted). We thus expect that such
systematic flow structures that depart from Keplerian rotation to be potentially detectable.
Moreover, some specific accreting layers in the disk surface, in both aligned and anti-aligned
cases, have accretion velocities near or exceed the sound speed. Note that these
layers are very thin and contain only a very small fraction of disk mass. Detecting such flow
structure would provide smoking-gun evidence of our simulation predictions, but it is also
very challenging, since it would require specific tracers whose optical depth $\tau\sim1$
surface is right in the vicinity of the thin accreting layer.

\section[]{Summary}\label{sec:sum}

In this work, we conducted the most comprehensive/realistic global simulations of the inner
regions of PPDs to date, that have incorporated all non-ideal MHD effects coupled to
steady-state chemistry with dust grains, as well as proper ray tracing schemes to calculate
disk ionization and control thermodynamics. All simulations include net poloidal magnetic
flux, which is an essential ingredient to launch MHD disk winds and drive disk accretion.
We have largely focused on a set of fiducial simulations, which give accretion rates on
the order of $10^{-8}M_{\bigodot}$ yr$^{-1}$, but also briefly explored the role of poloidal
field strength and FUV penetration depth. Our main findings from the fiducial simulations
are as follows.

\begin{itemize}
\item The bulk disk is largely laminar, launching an MHD disk wind that drives disk accretion.
The wind is magneto-thermal in nature, launched by magnetic pressure gradient, with very
strong mass loss rate that is comparable or larger than wind-driven accretion rate.

\item In the aligned case, the Hall shear instability (HSI) strongly amplifies horizontal field,
making the outcome dependent on initial field configuration. Mimicking realistic initial
conditions, we find that the disk is divided into an asymmetric inner zone
(within $\sim10$ AU) and a more
symmetric outer zone that are smoothly connected. Accretion and decretion flows at the
$\sim1\%$ of sound speed coexist in the bulk disk, determined by the vertical gradient
of toroidal field.

\item In the aligned case, both MHD wind and the laminar Maxwell stress contribute at
comparable level to disk accretion. However, since the latter contribution depends on
its radial gradient, making local accretion rates sensitive to radial disk structure, and
the accretion process is likely non-steady.

\item In the anti-aligned case, the disk may achieve a symmetric state that rapidly loses
magnetic flux, or (more likely) an asymmetric state that retain magnetic flux. In the latter
case, accretion is predominantly wind-driven, with the bulk accretion flow located at
one side of the disk surface at transsonic to supersonic speed. The resulting disk wind
is highly asymmetric, with most mass loss at the opposite side of the accretion flow.
\end{itemize}
In addition, increasing poloidal field strength enhances mass accretion rate more than
enhancing mass loss rate. By contrast, increasing FUV penetration enhances mass loss
rate more than accretion rate.
Moreover, the mass loss rates found in these additional simulations are sufficiently high
to substantially shield the incoming FUV radiation. We find wind launching can still
operate in the X-ray ionization dominated disk surface layer when FUV is largely shielded,
with gas marginally coupled to the magnetic field through ambipolar diffusion.

We raised several outstanding issues in PPD gas dynamics in Section \ref{ssec:issues},
and our global simulations have largely clarified the issues on wind kinematics and
symmetry.  Our simulations have also found complex behaviors in magnetic flux evolution,
but in general the disks appear to be able to retain magnetic flux, or lose flux much more
slowly than found in idealized simulations \citep{BaiStone17}. 

Our simulation results also have major implications on planet formation, especially
that the complex flow structure may transform our understandings on how solids are
transported in disks. It also calls for a reassessment of other stages of planet
formation and migration. Based on the results, we further speculate that the solar
nebula was originally threaded by poloidal fields aligned with disk rotation.

\subsection[]{Limitations and Future Directions}

Given the richness of the results from our fiducial simulations, we have only very briefly
explored the parameter space. We expect the simulation results to be representative,
and the physics we have explained to be widely applicable. Further parameter exploration
may lead to additional variations and complications. In particular, the gas dynamics of
the bulk disk can be affected by grain abundance, as well as ionization rates, especially
the X-ray properties of the protostar.

Several of our simulations (especially those with anti-aligned polarity and deeper FUV
penetration) show signs of turbulence. Moreover, the stability of the strong current
layer from the HSI in the aligned case also requires further investigation. Future 3D
simulations are necessary to properly characterize their properties.

In addition, we have only treated thermodynamics in the bulk disk and the wind zone
very roughly. Fully self-consistent calculations would require coupling radiative transfer
and chemistry (including photo-chemistry) with dynamics (as in some photo-evaporation
simulations, e.g., \citealp{Owen_etal10,WangGoodman17}). These
treatments are essential to better determine wind kinematics and compare with observations,
and call for future investigations.

Finally, we have only explored the inner region of the disk ($\sim0.6-20$ AU). Future
explorations should also focus on other radial ranges, including the innermost region
(e.g., \citealp{Flock_etal17}) where most exoplanets are found, and the outer disk regions
which are more accessible with spatially-resolved observations. Moreover, simulations
would also benefit from using further larger domain size to potentially capture the fast
magnetosonic points, and to cover broader dynamical ranges.

\acknowledgments

I thank the referee for a prompt and detailed report, and acknowledge support from Institute
for Theory and Computation, Harvard-Smithsonian Center for Astrophysics. Computations
for this work are performed on the Hydra cluster managed by the Smithsonian Institution, and
on Stampede at the Texas Advanced Computing Center through XSEDE grant TG-AST140001.

\appendix

\section[]{A. Dependence on Initial Condition in the Aligned Case}\label{app:init}

\begin{figure*}
    \centering
    \includegraphics[width=180mm]{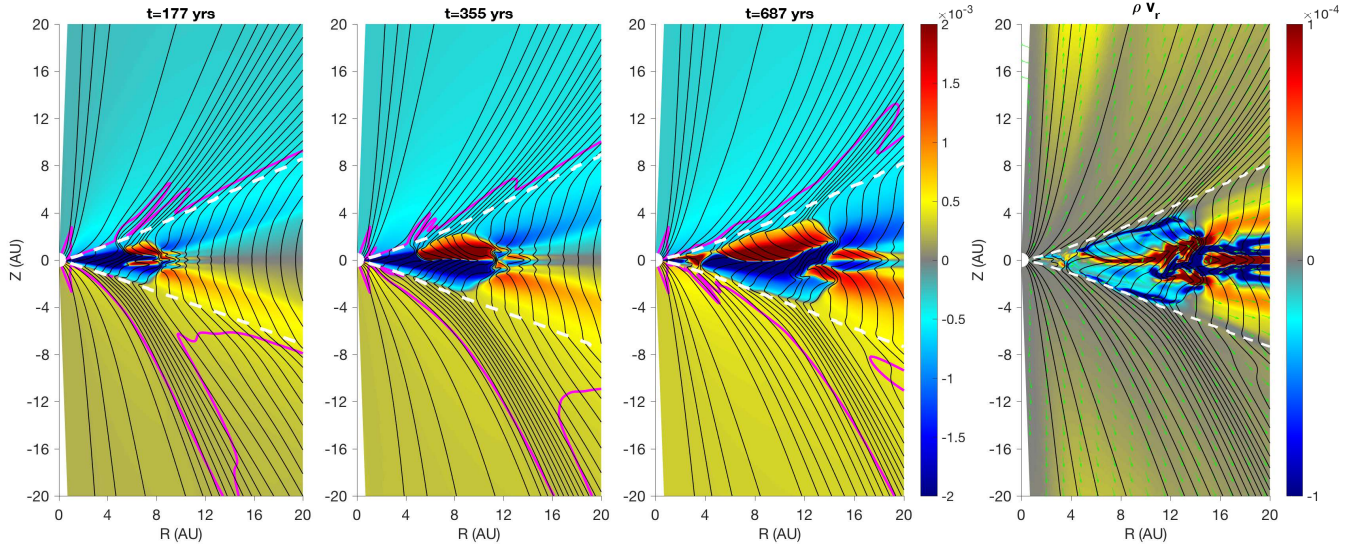}
  \caption{Same as Figure \ref{fig:hp_bevolve}, but all three non-ideal MHD terms are
  applied at the beginning of the simulation.}\label{fig:hp_bevolve_init}
\end{figure*}

In this appendix, we address the dependence on initial conditions in simulations with
all three non-ideal MHD effects included and initial vertical field aligned with disk rotation.
In Figure \ref{fig:hp_bevolve_init}, we show the time evolution of magnetic field
configuration from a modified version of Fid$+$, where all three non-ideal MHD terms are
turned on from the beginning (instead of turning on the Hall term from inside out).
In other words, the setup is the same as run Fid$-$ except that field polarity is flipped.

We immediately notice that during the evolution, there are patches in the disk that
possess strong toroidal field with opposite and alternating signs. The presence of
these patches is a result of the Hall-shear instability (HSI), as discussed in
\citet{Kunz08,Lesur_etal14} and \citet{Bai14} via local analysis and simulations.
The global manifestation of the HSI was studied in \citet{BaiStone17}, and is briefly
reviewed in Section \ref{ssec:hp_bevolve}.

The nature of the HSI dictates that the formation of such discrete patches depends
on initial condition, and can be stochastic. For instance, a random perturbation in
vertical field can create regions of radial field with opposite signs, with each region
growing their own HSI, eventually creating a pair of relatively strongly magnetized
patches.
Indeed, we have also run simulations with different field geometries
(controlled by the $m$ parameter in Equation (\ref{eq:Aphi})), and found qualitatively
similar outcomes except that these patches are distributed differently.

The nature of the HSI also dictates that these strongly magnetized patches (but still
not sufficiently strong to become magnetically dominated), once formed, do not
annihilate with neighboring patches with opposite sign of toroidal field. Instead,
each patch attempts to expand its ``territory" through the growth of HSI. At the
boundaries, the balance between growth and resistivie/ambipolar dissipation
allows these patches to survive and evolve slowly. Because the flow structure in
the disk strongly depends on the vertical gradient of toroidal field described by
Equation (\ref{eq:radflow}), we see in the rightmost panel of Figure
\ref{fig:hp_bevolve_init} that strong radial flows are induced at patch boundaries.

The sensitive dependence on the initial conditions may suggest that in the aligned
case, the overall magnetic field structure, and hence the flow structure in the
disk is somewhat unpredictable. While this is entirely possible and worth further
investigations, it also suggests that we need to consider setting up the initial conditions
in a more realistic way that mimics the initial stage of disk formation.

Formation of PPDs follows from protostellar core collapse, and it is well known that
non-ideal MHD effects play an important role throughout the processes of
core collapse and disk formation (see \citealp{Li_etal14} for a review, and more recent
works by \citealp{Tomida_etal15,Tsukamoto_etal15,Wurster_etal16}).
We note that these processes are accompanied by rapid increase of gas density,
and the development of more rapid rotation that winds up poloidal field into toroidal
field. According to the relative ordering among the three non-ideal MHD effects,
AD is the dominant effect at the beginning, the Hall term gradually picks up as density
increases. The key notion here is that before the Hall effect becomes important, the
system has already evolved under AD with well-defined sense of (differential) rotation
to produce toroidal field with ordered vertical structures. As density builds up further,
the Hall effect enters and modifies the field structure through the HSI.

The discussion above motivates us to adopt the procedure described in Section
\ref{sec:align} to set up the initial conditions. The key is to allow the system to
develop some initial toroidal field structures
before introducing the Hall effect, so that the HSI can grow on top of these pre-existing
structures. This procedure, when applied to run Fid$+$, is very successful in minimizing
the number of discrete magnetized patches to result in sufficiently simple field structure,
which also appears more physically reasonable.

However, the same procedure appears less successful for some other runs in our
parameter study (Section \ref{sec:param}). This again reflects the sensitive dependence
of the outcome on initial conditions, but it is also true that our new procedure of setting up
the initial condition is still far from representing realistic conditions of disk formation.
In view of these results, we are still argue for the generality of the field structure obtained
in run Fid$+$, although future works are needed to reach firm conclusions.

\section[]{B. Further Discussions on the Symmetry Issue in the Aligned Case}\label{app:symmetry}

Our simulation results resolve a number of puzzles found in previous local simulations.
First, previous local shearing-box simulations of \citet{Lesur_etal14} and \citet{Bai14}
always found that in the inner disk, toroidal field of a single sign overwhelms the entire
box after the development of the HSI even the vertical box extends into the FUV layer.
The failure for $B_\phi$ (and hence $B_R$) to flip across the disk means that while the
system launches an outflow, there is no net transport of angular momentum. Our results
show that toroidal field does flip, and a fully global setup is essential which avoids the
influence of artificial boundary conditions in local simulations.

Second, local simulations of \citet{Bai14,Bai15} also found that towards outer radii
($\gtrsim5$ AU), flipping the toroidal field within the simulation box is possible, but it
was unclear how this field configuration connects to the inner radii. Our simulations
demonstrate that they join smoothly. 

In brief, the inner zone with strong asymmetry is related to strong Hall effect.
Development of such asymmetry is unavoidable in these simulations because
within a few AU, resistivity is so strong that a flipped field configuration is
not sustainable. This is the reason why simulations of \citet{BaiStone17} also
dominated by the Hall effect yet does not produce the asymmetric field configuration.

\section[]{C. Nomenclature of PPD Disk Winds}\label{app:interpwind}

In this appendix, we aim to clarify the nomenclature of disk winds from PPDs, in
light of recent development of magnetized wind theory and global disk simulations.
Broadly speaking, disk winds can be thermally-driven (by thermal pressure),
magnetically-driven (by the Lorentz force), or driven by radiation pressure (e.g.,
\citealp{Proga07}). For PPDs in T-Tauri phase, radiation pressure
is largely negligible due to their low luminosities (e.g., \citealp{Cabrit07}).
We focus on thermal and magnetic effects.

Thermally-driven PPD wind is generally referred to as {\it photoevaporation}. It
results from external heating by high-energy photons (UV to X-ray), either from the
central protostar, or nearby massive stars. Extensive literature has focused on
this type of disk winds (see \citealp{Alexander_etal14} for a review), and the
outcome is very sensitive to details of the heating and cooling processes.

Magnetically-driven wind has two flavors.
\begin{itemize}
\item {\it Magnetocentrifugal wind}: strong poloidal field lines anchored to the disk
enforces the outflow to corotate with the wind foot-point (wind base), leading to
centrifugal acceleration as viewed from the corotating frame when field geometry is
favorable. It is directly related to the \citet{BlandfordPayne82} scenario.
\item {\it Wind driven by magnetic pressure gradient}: poloidal field
lines are too weak to enforce corotation, and get winded up to build up toroidal
field. The wind is launched by vertical gradient of magnetic pressure from the
toroidal field. It is sometimes referred to as magnetic tower flows
\citep{LyndenBell96,LyndenBell03}.
\end{itemize}
Viewed in the observer's frame, the Lorentz force is the driving force in both cases,
again from the pressure gradient and tension force from the toroidal field
\citep{Spruit96}. Nevertheless, it is physically intuitive to distinguish the
two regimes, as is widely adopted in other contexts such as star formation
(e.g., \citealp{Seifried_etal12,Tomida_etal13}). The centrifugally-driven wind
typically corresponds to large Alfv\'en radius (long lever arm), and the wind is
lightly loaded. The opposite regime corresponds to small Alfv\'ven radius (short lever
arm), with a heavily loaded wind. Fixing other conditions, one generally smoothly
transitions from magnetic pressure gradient driven wind to centrifugally driven wind
as poloidal field strength increases \citep{Bai_etal16}.

Conventionally, magnetically-driven wind models/simulations usually assume a cold gas
flow and hence thermal pressure plays a negligible role throughout the process
(with a few exceptions, such as \citealp{CasseFerreira00b}). 
In principle, both magnetic and thermal effects can contribute to the launching and
acceleration the wind flow. 

We call a disk wind ``magneto-thermal" when the wind properties are affected both
the field strength and thermodynamics. This term can be considered to be broadly defined,
encompassing the aforementioned wind driving mechanisms as long as both magnetic and
thermal effects play a role.
One can imagine that by applying poloidal fields to a pure thermal disk wind and increasing
the field strength, the wind will transition to being driven by magnetic pressure gradient,
and eventually become centrifugally-driven.
In general, PPD winds are magneto-thermal, because poloidal magnetic fields are essential
to drive disk accretion, and strong external heating is inevitable that can drive a thermal wind
on its own. We have shown in \citet{Bai_etal16} as well as in this paper that magnetic
pressure gradient is the main wind launching mechanism.

Many examples of magneto-thermal winds are explored in idealized models of \citet{Bai_etal16},
where the wind is assumed to be launched in the ideal MHD regime from the wind base at the
disk surface that is well separated from the poorly ionized disk main body, with a barotropic
equation of state. The reality can always be more complex, as some of our simulations and
the ones in BLF17 illustrate. Here we list two scenarios where wind properties are strongly
modified by thermal effects, which can be considered as key characteristics of magneto-thermal
disk winds:
\begin{itemize}
\item Wind launching takes place where magnetic pressure is not much larger than thermal
pressure, as studied in \citet{Bai_etal16}. The resulting wind is typically very heavily
loaded, with very small lever arm. Sometimes it even violates the requirement that the
lever arm $\lambda>3/2$ for a cold MHD wind \citep{CasseFerreira00b}, as we have observed
in several cases in our fiducial simulations, as well as in BLF17.
\item The wind is launched magnetically, but is subsequently accelerated by thermal
pressure from external heating. This case includes the hot wind simulations in BLF17. We
studied a class of models of this type in \citet{Bai_etal16}, finding that despite the
subsequent wind acceleration from external heating, wind lever arm and hence wind mass
loss rate is largely unaffected. This is also confirmed in the BLF17 simulations.
\end{itemize}
Note that the above two scenarios do not necessarily exclude each other. In addition,
we have also found that when FUV is shielded by the wind itself (as we find in the case
of deeper FUV penetration and stronger magnetization), the wind can be launched,
and achieve super-Alfv\'enic velocities within the non-ideal MHD layer. Similar results
are found in BLF17. These can be considered as extensions of the first scenario, though
gaining more quantitative understandings is much less straightforward.

Finally, the terms ``MHD disk wind", or ``magnetized disk wind", generally refer to disk
winds that are launched magnetically without specifically referring to thermal
effects. We consider these terms to be more broad and inclusive, and can be applied to
magnetized PPD winds in general.

\bibliographystyle{apj}

\label{lastpage}
\end{document}